\documentclass[final,1p,times]{elsarticle}

\usepackage{amssymb,amsmath,amsfonts}
\usepackage[graphicx]{realboxes}
\usepackage{lineno} 
\usepackage{multirow}
\usepackage{pdflscape} 
\usepackage{hyperref}
\hypersetup{
    colorlinks=true,
    linkcolor=blue,
    filecolor=magenta,      
    urlcolor=cyan,
    pdftitle={On Xmax statistics},
    pdfpagemode=FullScreen,
    }
    
\newcommand{\epos}{\texttt{EPOS LHC-R}}
\newcommand{\sibyll}{\texttt{Sibyll 2.3e}}
\newcommand{\qgsjet}{\texttt{QGSJet III-01}}
\newcommand{\conex}{\texttt{CONEX}}

\graphicspath{{figs/}}

\journal{Astroparticle Physics}

\begin{document}

\begin{frontmatter}



\title{Updated Air-Shower \( X_{\rm max} \) Moment Parametrizations for UHECR Composition with Latest Hadronic Interaction Models}
\author[gssi,infn]{Carmelo Evoli}
\author[gssi,infn]{Igor Vaiman}
\author[gssi,infn]{Sergio Petrera}
\author[univaq,infn]{Francesco Salamida}
\affiliation[gssi]{organization={Gran Sasso Science Institute (GSSI)},
            addressline={Viale Francesco Crispi 7}, 
            city={L'Aquila},
            postcode={67100}, 
            state={AQ},
            country={Italy}}
\affiliation[univaq]{Organization={Università degli Studi dell’Aquila, Dipartimento di Scienze fisiche e chimiche},
            addressline={Via Vetoio}, 
            city={L'Aquila},
            postcode={67100},
            state={AQ},
            country={Italy}}
\affiliation[infn]{organization={INFN-Laboratori Nazionali del Gran Sasso(LNGS)},
            addressline={Via G.~Acitelli 22}, 
            city={Assergi},
            postcode={67100}, 
            state={AQ},
            country={Italy}}
            
\begin{abstract}
The mass composition of ultra-high-energy cosmic rays (UHECRs) is commonly inferred from the first two moments of the depth of shower maximum, $X_{\rm max}$, measured by fluorescence and hybrid detectors. Such analyses require fast and accurate mappings between the moments of $X_{\rm max}$ and those of the logarithmic mass, $\ln A$, based on realistic air-shower simulations. In this work we provide updated parametrizations of the $X_{\rm max}$ moments and distributions for air showers initiated by nuclei from proton to iron, simulated with \conex\ for three state-of-the-art hadronic interaction models: \textsc{Epos-LHC}, \textsc{Sibyll,2.3e}, and \textsc{QGSJet-III-01}.

We parametrize the mean depth $\langle X_{\rm max}\rangle$ and the variance $\sigma^2(X_{\rm max})$ as functions of energy and mass. For the variance we compare a second-order polynomial model with an exponential model. 
In addition, we model the full $X_{\rm max}$ distributions with a three-parameter generalized Gumbel function. The Gumbel parameters are fitted using an unbinned likelihood and are validated by comparing the implied mean and variance with the raw \conex\ samples and with the moment parametrizations.

Across the full energy range considered, residuals between the parametrizations (or the Gumbel representation) and the simulations are at the level of a few g~cm$^{-2}$ for the mean and a few (g~cm$^{-2}$)$^2$ for the variance, making these parametrizations suitable for precision UHECR composition studies and forward-folding analyses of $X_{\rm max}$ distributions.
\end{abstract}


\end{frontmatter}



\section{Introduction}
\label{sec:intro}

The mass composition of ultra-high-energy cosmic rays (UHECRs) remains one of the central unknowns in astroparticle physics, with direct implications for the identification of their sources, acceleration mechanisms, and propagation effects. Among the shower observables accessible to current fluorescence and hybrid detectors, the most widely used composition-sensitive quantities are the mean depth of shower maximum, $\langle X_{\rm max}\rangle$, and its fluctuations, commonly expressed through the dispersion $\sigma(X_{\rm max})$. These observables provide a compact description of the longitudinal development of extensive air showers and can be related, in a statistical sense, to the mass and energy of the primary particle~\cite{Kampert:2012mx}.
	
Interpreting $\langle X_{\rm max}\rangle$ and $\sigma(X_{\rm max})$ in terms of primary mass, however, unavoidably relies on air-shower simulations, and thus on the adopted hadronic interaction model (HIM). Modern HIMs are all tuned to reproduce collider data at accessible energies, but they differ in the physical assumptions used to extrapolate multiparticle production, cross sections, and secondary spectra into the kinematic regime relevant for UHECR-induced interactions. As a consequence, the predicted $\langle X_{\rm max}\rangle$ and $\sigma(X_{\rm max})$ for a given primary mass can differ significantly among models~\cite{PierreAuger:2013xim}. Despite substantial progress in hadronic interaction modeling at ultra-high energies, air-shower physics beyond terrestrial accelerators still carries sizable and difficult-to-quantify systematic uncertainties (see, e.g., Ref.~\cite{Albrecht:2025kbb}), which propagate directly into composition inferences.
The question of how to use $X_{\rm max}$ information to constrain composition in the presence of hadronic-model uncertainties has motivated several dedicated studies, including early work based on moment parametrizations and their application to experimental data~\cite{PierreAuger:2013xim}. 

Over the past decade, continuous improvements in fluorescence reconstruction and atmospheric monitoring have enabled increasingly precise measurements of $X_{\rm max}$. The Pierre Auger Observatory reports systematic uncertainties on $X_{\rm max}$ of order $10~{\rm g\,cm^{-2}}$~\cite{PierreAuger:2024nzw}, while the Telescope Array quotes $\pm 17~{\rm g\,cm^{-2}}$~\cite{Bergman:2020izr}. 
That is why, in parallel with these experimental advances, recent composition studies increasingly exploit the \emph{full} $X_{\rm max}$ distribution rather than only its first two moments~\cite{PierreAuger:2014gko}. This shift places stringent requirements on the modeling of the distribution shape, because several steps in the analysis chain depend on an a priori description of the expected $X_{\rm max}$ probability density. A widely adopted approach is to represent simulated $X_{\rm max}$ distributions with Gumbel-type functions~(see~\cite{DeDomenico:2013wwa} and references therein), which have been shown to reproduce Monte Carlo predictions with high fidelity across relevant energy ranges and primary masses~\cite{Arbeletche:2019qma}. 

Reliable parametrizations of both the moments and the distribution shape therefore constitute an essential interface between air-shower simulations and experimental composition analyses.

The purpose of this paper is to provide an updated and self-consistent set of parametrizations for $\langle X_{\rm max}\rangle$, $\sigma(X_{\rm max})$, and the corresponding Gumbel description of the $X_{\rm max}$ distribution, based on the latest developments in hadronic interaction modeling. By fitting the parametrizations to current HIM implementations, we aim to deliver a practical tool for ongoing and forthcoming UHECR composition studies, facilitating comparisons between experiments and improving the robustness of mass-inference procedures against model-dependent effects.
We also compare $X_{\rm max}$ observables' parametrization between latest and previous versions of each HIM, quantifying the impact of the update on observables and highlighting energy ranges where it is the largest.

\section{CONEX simulations}
\label{sec:conex}

\begin{figure*}[t]
\centering
\hspace{\stretch{1}}
\includegraphics[width=0.40\textwidth]{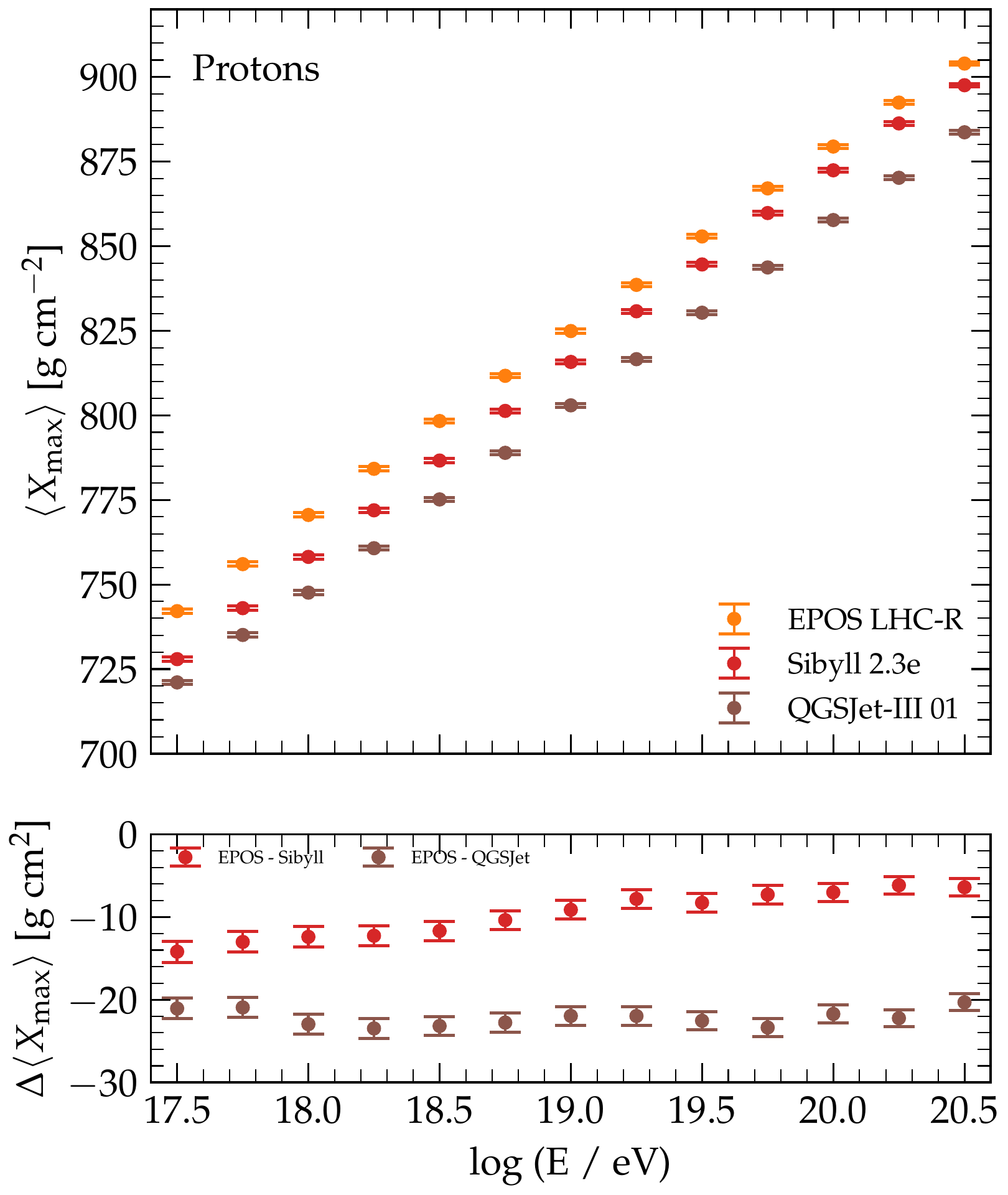}
\hspace{\stretch{1}}
\includegraphics[width=0.40\textwidth]{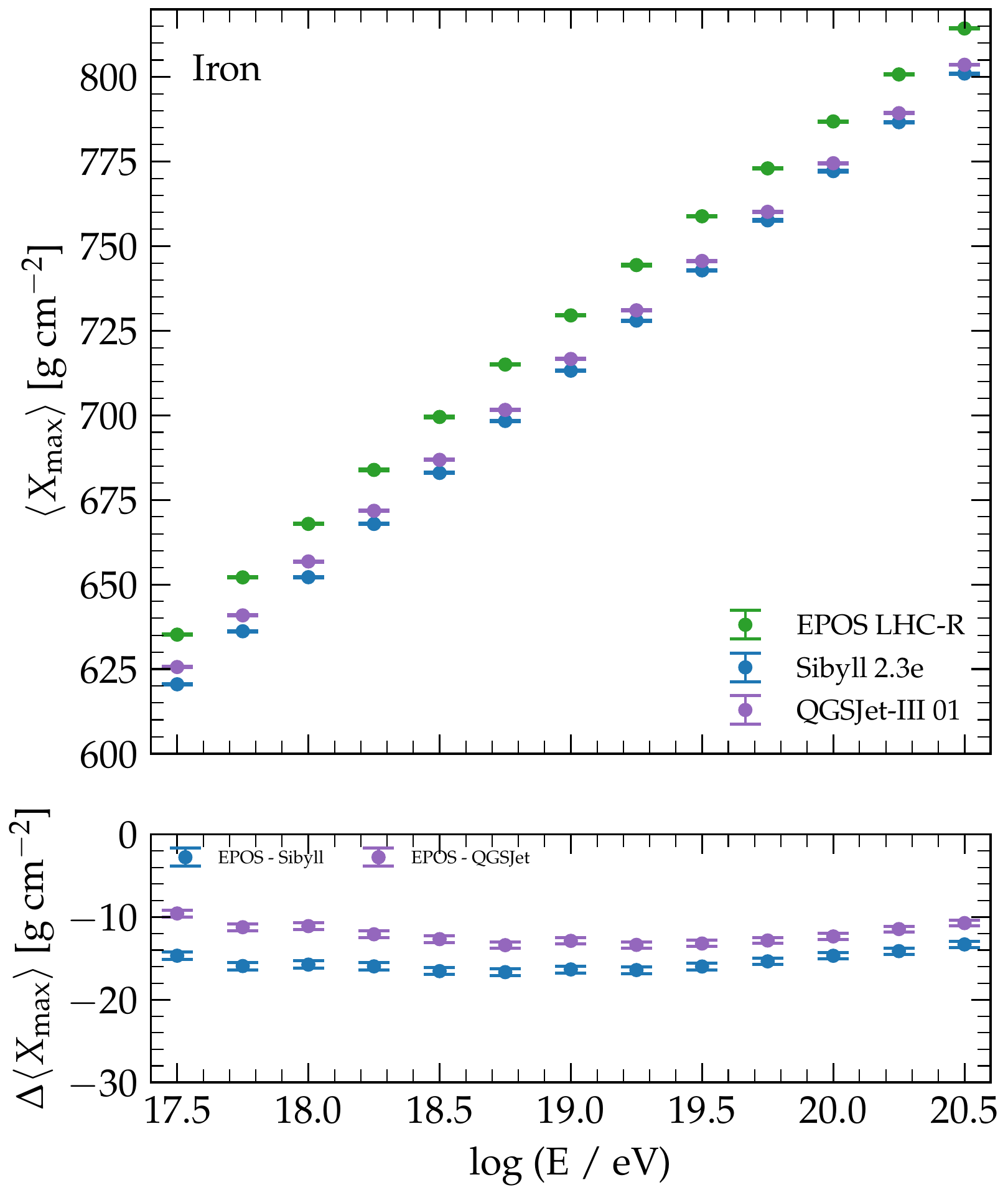}
\hspace{\stretch{1}}
\caption{Mean $X_{\mathrm{max}}$ as a function of energy for protons (left) and iron (right), 
as obtained for the three hadronic interaction models: \sibyll, \epos, and~\qgsjet. 
The lower panels display the systematic differences of \sibyll~and \qgsjet~with respect to \epos.}
\label{fig:xmax_mean}
\end{figure*}

\begin{figure*}[t]
 \centering
\hspace{\stretch{1}}
\includegraphics[width=0.40\textwidth]{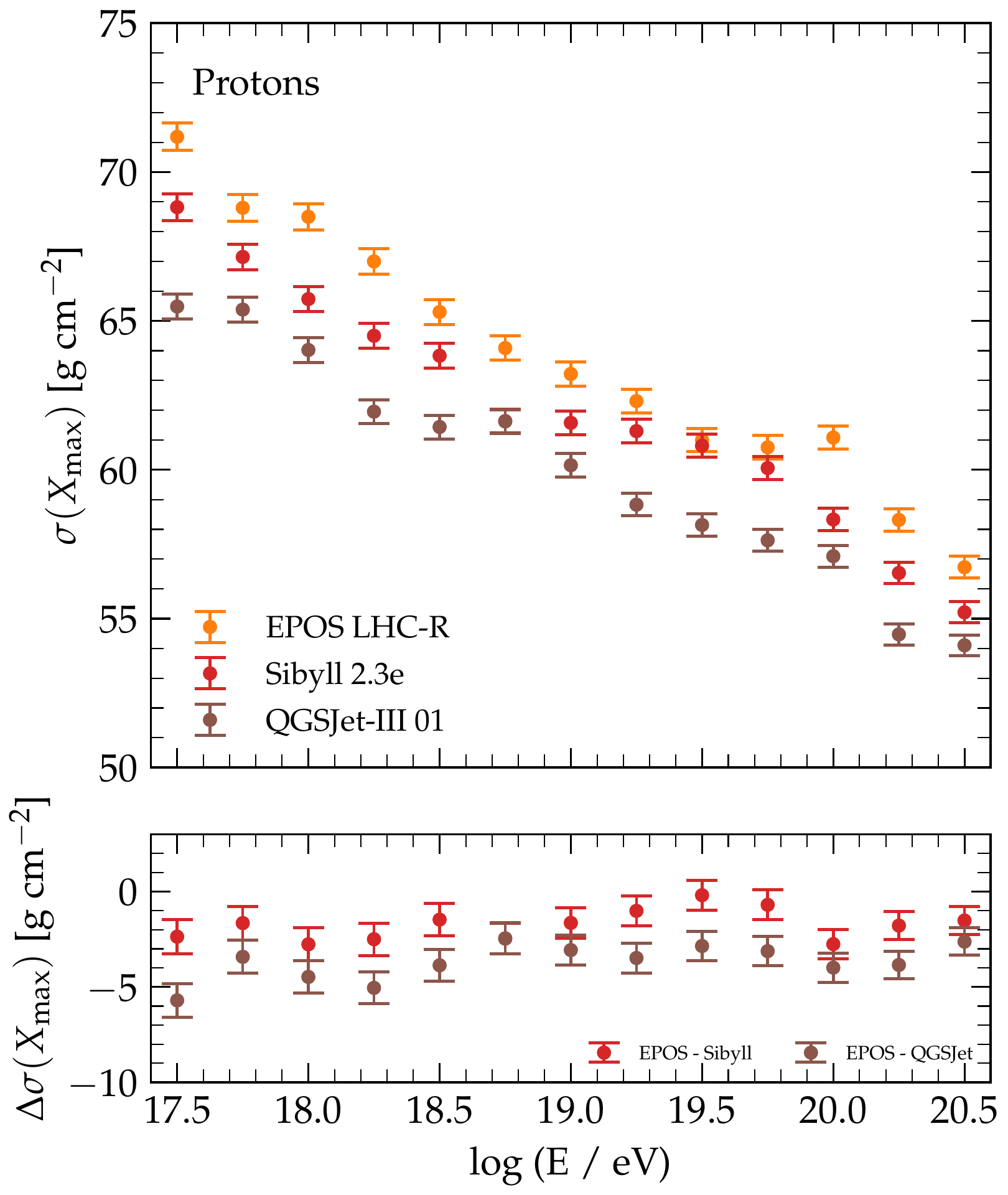}
\hspace{\stretch{1}}
\includegraphics[width=0.40\textwidth]{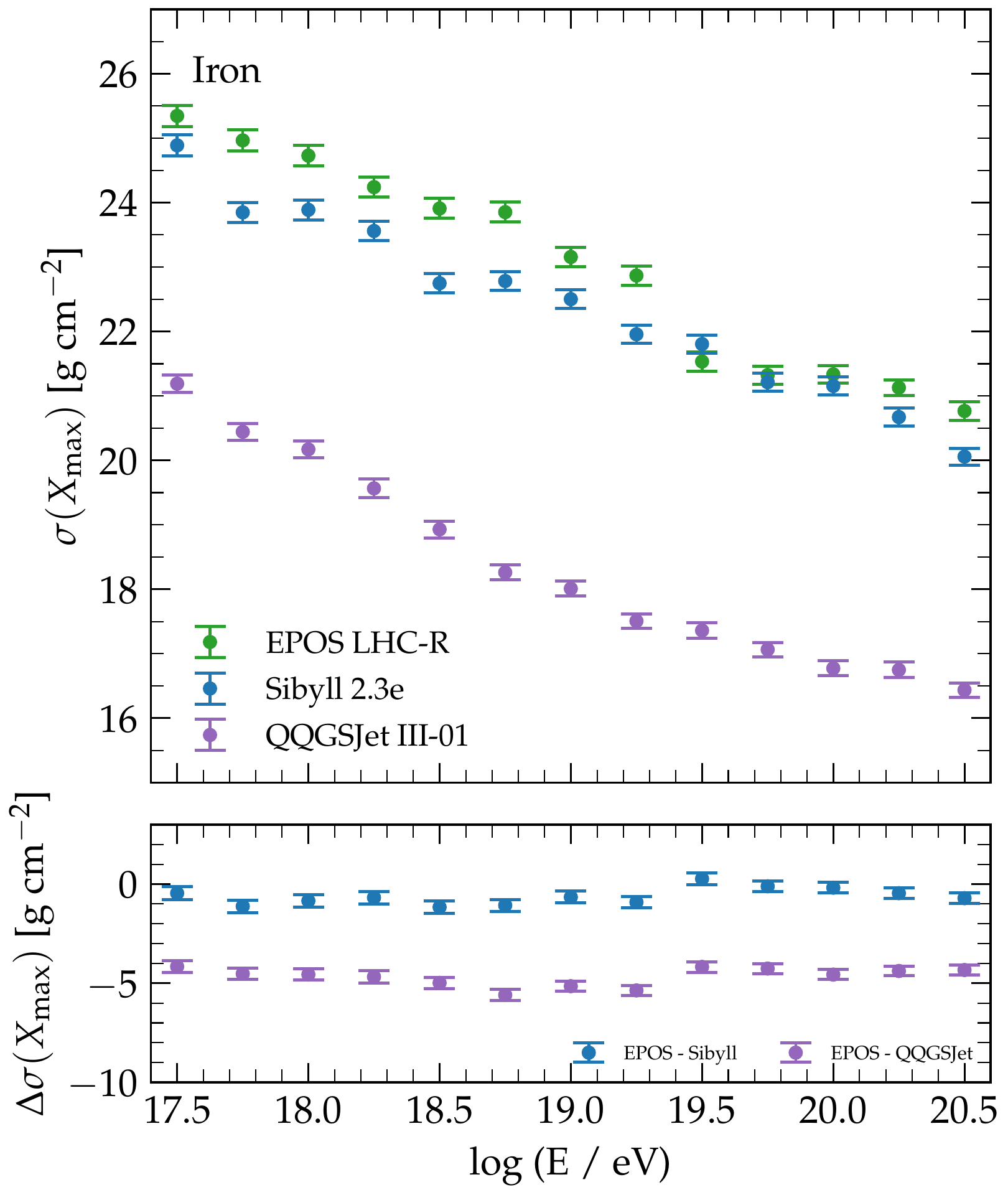}
\hspace{\stretch{1}}
\caption{Standard deviation of $X_{\mathrm{max}}$ as a function of energy for protons (left) and iron (right), as obtained for the three hadronic interaction models: \sibyll, \epos, and~\qgsjet. 
The lower panels display the systematic differences of \sibyll~and \qgsjet~with respect to \epos.}
\label{fig:xmax_sigma}
\end{figure*}

Dedicated \conex~simulations of extensive air-shower development were performed using version~7.801\footnote{\url{https://gitlab.iap.kit.edu/AirShowerPhysics/cxroot/-/tree/conex2r7.801}}. 
The $X_{\mathrm{max}}$ distributions used for the fits were obtained for the following hadronic interaction models:
\sibyll~\cite{Riehn:2019jet}, \epos~\cite{Pierog:2023ahq}, and \qgsjet~\cite{Ostapchenko:2024myl} (see~\ref{sec:oldhim} for a comparison with previous generations of these models).
Simulations were carried out for primary nuclei H ($A = 1$), He ($A = 4$), N ($A = 14$), Si ($A = 28$), and Fe ($A = 56$). 
The datasets cover the energy range $\log_{10}(E/\mathrm{eV}) = 17.5$--$20.5$, chosen to match the range used in mass-composition analyses of the Pierre Auger Observatory~\cite{PierreAuger:2014gko}. 

An extension of the simulations and parametrizations to heavier primaries and lower-energies is planned for a forthcoming paper.

Simulations are performed for 13 values of energy with step $\Delta \log_{10} E = 0.25$. For each energy, at least $10^4$ showers were simulated. The zenith angle was fixed at $60^\circ$, and no significant dependence of the results on this choice was observed.

The $X_{\mathrm{max}}$ values are taken from the \texttt{XmxdEdX} branch of the \conex~output, which corresponds to a quadratic interpolation of the three highest $dE/dX(X_{\rm max})$ points.

For each energy bin we computed the mean and variance of $X_{\mathrm{max}}$ over $N$ simulated showers as follows:
\begin{itemize}
\item Mean and error of the mean:
\begin{equation}
\mu_X = \frac{1}{N} \sum_{i=1}^N X_i \, , 
\qquad 
\delta \mu_X  = \frac{\sigma_X}{\sqrt{N}} \, ,
\end{equation}

\item Variance and error of the variance:
\begin{equation}
\sigma_X^2 = \frac{1}{N-1} \sum_{i=1}^N (X_i - \langle X \rangle)^2 \, , 
\qquad 
\delta \sigma_X^2 = \sigma_X^2 \sqrt{\frac{2}{N-1}} \, .
\end{equation}  
\end{itemize}

We present comparative plots of the mean $\mu_X = \langle X_{\mathrm{max}} \rangle$ and standard deviation $\sigma_X = \sqrt{\sigma^2({X_\mathrm{max}})}$ for the three hadronic interaction models, focusing on the two extreme cases of primary mass: protons and iron nuclei (Figs.~\ref{fig:xmax_mean} and~\ref{fig:xmax_sigma}, respectively). 

Across the full energy range considered, \epos~predicts the deepest shower maxima for protons, followed by \sibyll~and \qgsjet. For iron primaries, this ordering changes: \sibyll~yields the shallowest $X_{\mathrm{max}}$, while \epos~and \qgsjet~predict deeper shower maxima. 

For the fluctuations, significant differences also emerge between models. In general, \epos~predicts larger variances of $X_{\mathrm{max}}$, while \qgsjet~and \sibyll~lead to narrower distributions, especially for proton primaries. In the case of iron, the \qgsjet~predictions deviate significantly from those of the other two models, with direct implications for mass-composition inferences based on this model. 
Such comparisons are particularly relevant for assessing the systematic uncertainties associated with the choice of hadronic interaction model in mass-composition analyses.

\section{$X_{\mathrm{max}}$ parametrizations of mean and variance above $10^{17.5}$ eV}
\label{sec:xmax}

In this section we describe the functional forms adopted to parametrize the mean and variance of the $X_{\mathrm{max}}$ distributions.  The parametrizations are conveniently expressed in terms of two variables:
\begin{equation}
\epsilon \equiv \log_{10}(E/E_0) \, , 
\qquad 
Y \equiv \ln A \, ,
\end{equation}
where $E$ is the primary energy, $E_0 = 10^{19}~\mathrm{eV}$ is a reference energy, and $A$ is the mass number of the primary nucleus. To simplify notation, $X_{\mathrm{max}}$ is also abbreviated as $X$ throughout.

In the context of the so-called {\it superposition model} the mean depth of maximum is a function of the energy-per-nucleon, $E/A$. The dependence is commonly assumed to be linear in its logarithm, following the Heitler model~\cite{Matthews:2005sd}. In this paper we use a generalization of these models and, following~\cite{PierreAuger:2013xim}, we use a parametrization with four free parameters ($a_0$, $a_1$, $b_0$, $b_1$):
\begin{eqnarray}
a^\prime(\epsilon) & = & a_0 + a_1 \, \epsilon \nonumber , \\ 
b^\prime(\epsilon) & = & b_0 + b_1 \, \epsilon \nonumber , \\ 
\mu_X (\epsilon, Y) & = & a^\prime(\epsilon) + b^\prime(\epsilon)\, Y \, .
\label{eq:xmax_mean}
\end{eqnarray}
This expression separates the pure energy dependence, $a^\prime(\epsilon)$, from the linear mass dependence encoded in $b^\prime(\epsilon)\,Y$.

For the variance of the distributions we consider two alternative parametrizations. The first follows~\cite{PierreAuger:2013xim} and is based on a second-order polynomial dependence on $\epsilon$ and a quadratic expansion in $Y$, for a total of six free parameters ($c_0$, $c_1$, $c_2$, $d_0$, $d_1$, $d_2$):
\begin{eqnarray}
c^\prime(\epsilon) & = & c_0 + c_1 \, \epsilon + c_2 \, \epsilon^2 \nonumber  ,\, \\
d^\prime(\epsilon) & = & d_0 + d_1 \, \epsilon \nonumber  ,\, \\
\sigma_X^2 (\epsilon, Y) &=& c^\prime(\epsilon) \left[ 1 + d^\prime(\epsilon)\, Y + d_2 \, Y^2 \right] \, .
\label{eq:xmax_sigma2}
\end{eqnarray}

The second parametrization is based on an exponential dependence on $Y$:
\begin{eqnarray}
f^\prime(\epsilon) & = & f_0 + f_1 \, \epsilon \, , \nonumber \\
g^\prime(\epsilon) & = & g_0 + g_1 \, \epsilon \nonumber, \\
\sigma_X^2(\epsilon, Y) & = & \exp\!\left[ f^\prime(\epsilon) + g^\prime(\epsilon)\, Y \right] \, .
\label{eq:xmax_sigma2_exp}
\end{eqnarray}
This form guarantees $\sigma_X^2 > 0$ for all $\epsilon$ and $Y$ and introduces four free parameters ($f_0$, $f_1$, $g_0$, $g_1$).

The first parametrization, Eq.~\eqref{eq:xmax_sigma2}, has the advantage of leading to a simple analytical inversion when mapping $\sigma^2_{X_{\mathrm{max}}}$ to $\sigma^2_{\ln A}$ in composition analyses. However, its polynomial dependence on $Y$ can lead to an unphysical upturn of $\sigma_X^2$ outside the validity range in mass, due to the quadratic term in $Y$.

The exponential parametrization, Eq.~\eqref{eq:xmax_sigma2_exp}, typically yields smaller residuals with respect to the simulated variances and can be more robust when extrapolated to higher masses, even though the inversion from $\sigma^2_{X_{\mathrm{max}}}$ to $\sigma^2_{\ln A}$ is less simple (see also~\cite{Mollerach:2025fmx}).
All these aspects will be treated in more details in Sec.~\ref{sec:lnainversion}.

\subsection{Fit parameters}

The best-fit values of the parameters for each hadronic interaction model are reported in Table~\ref{tab:xmax_moments_params} for $\mu_X$ and $\sigma^2_X$, respectively. These values were obtained by fitting the statistics of the \conex~simulation datasets introduced in Section~\ref{sec:conex}.

The fits were performed with the \texttt{iminuit}~\citep{iminuit} minimization package, using the standard \texttt{MIGRAD} $\chi^2$ minimization. The reduced $\chi^2$ values ($\chi^2/\mathrm{ndf}$) corresponding to the best-fit parameters are listed in the last column of the table. 

The $\chi^2$ is computed using, as uncertainties, the statistical errors on the mean and variance as defined in Section~\ref{sec:conex}. For this reason, the absolute $\chi^2$ values should not be interpreted as a measure of the overall goodness-of-fit, but rather as a relative indicator to compare the quality of the fits among different hadronic interaction models.

\begin{figure}[t]
\centering
\includegraphics[width=0.32\textwidth]{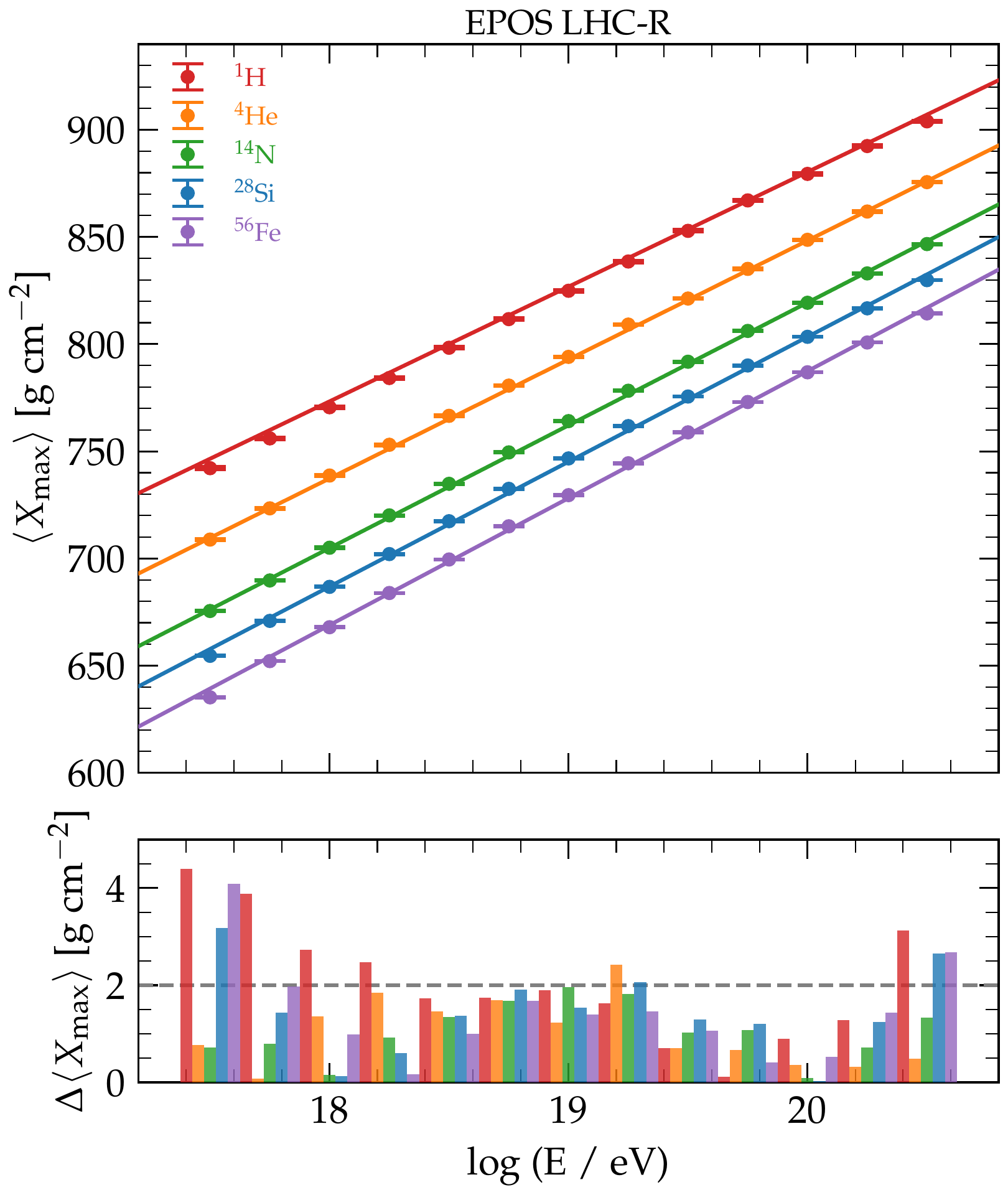}
\includegraphics[width=0.32\textwidth]{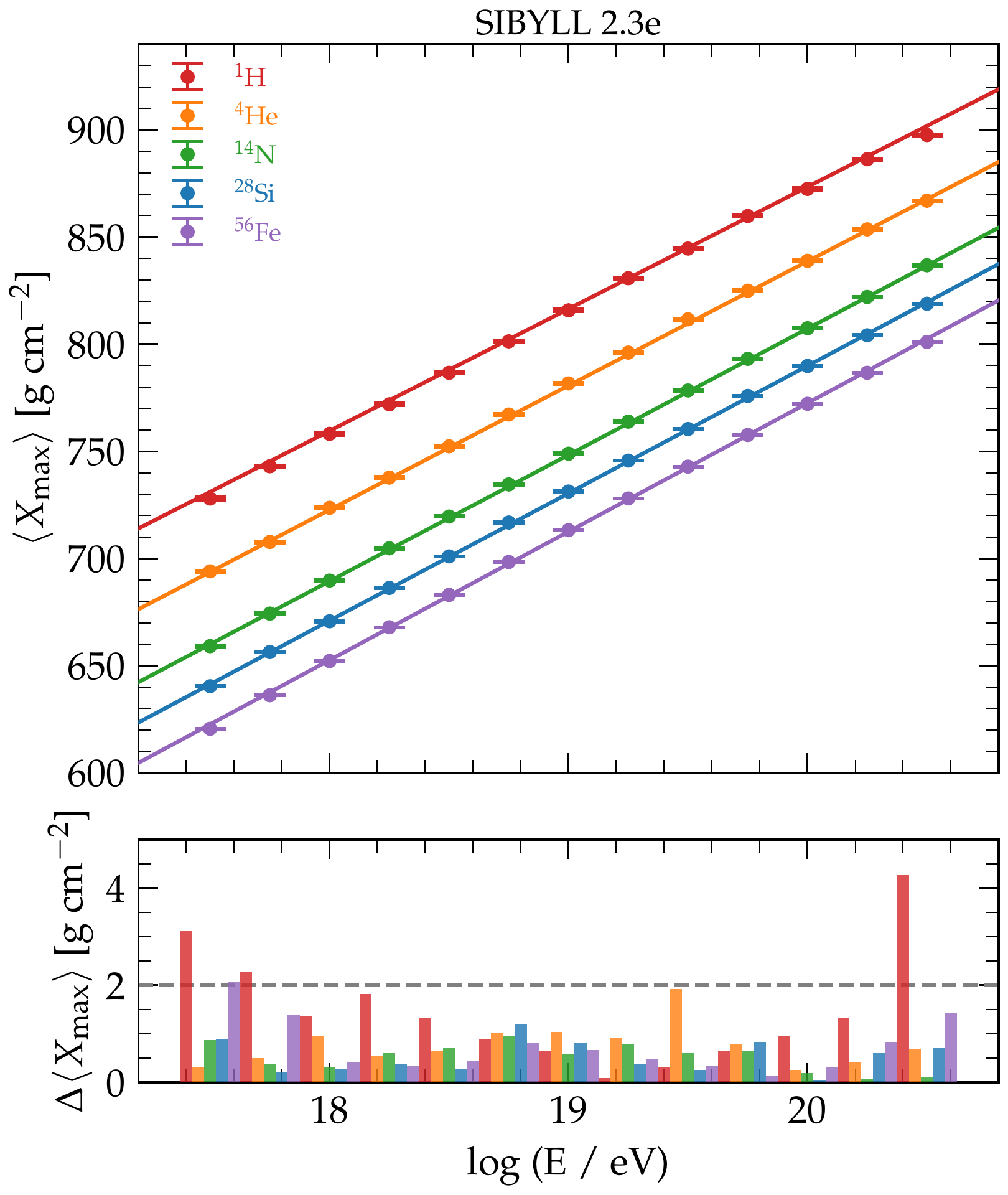}
\includegraphics[width=0.32\textwidth]{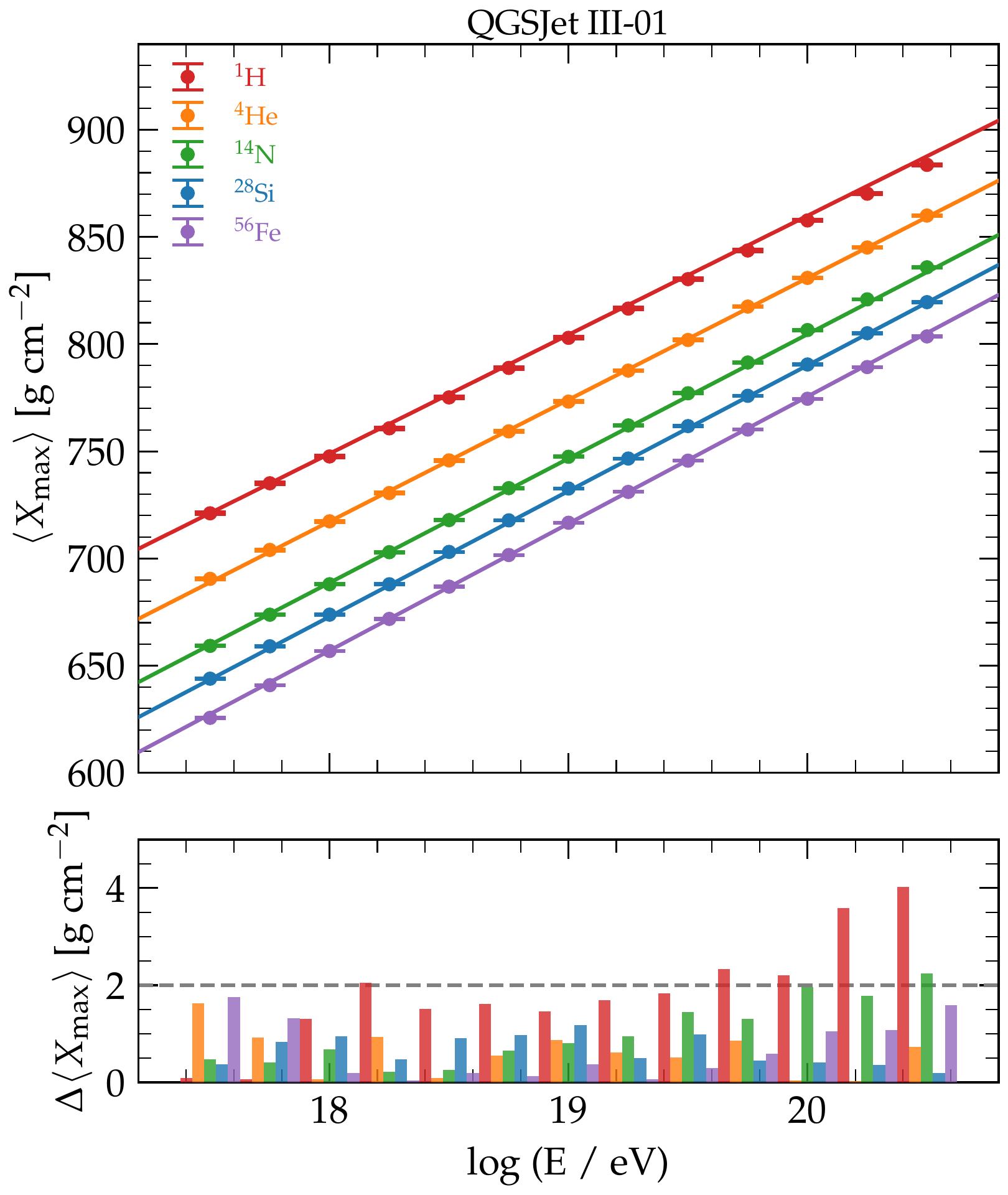}
\caption{Comparison between the parametrizations and the \conex~simulations for 
$\langle X_{\mathrm{max}} \rangle$ for the three hadronic interaction models: \epos\ (left), 
\sibyll\ (middle), and \qgsjet\ (right). The upper panels show the simulated values (dots) 
together with the analytical parametrizations (solid lines) for representative primary nuclei. 
The lower panels display the residuals, defined as in Eq.~\ref{eq:bias}.}
\label{fig:xmax_residual}
\end{figure}

\begin{figure}[t]
\centering
\includegraphics[width=0.32\textwidth]{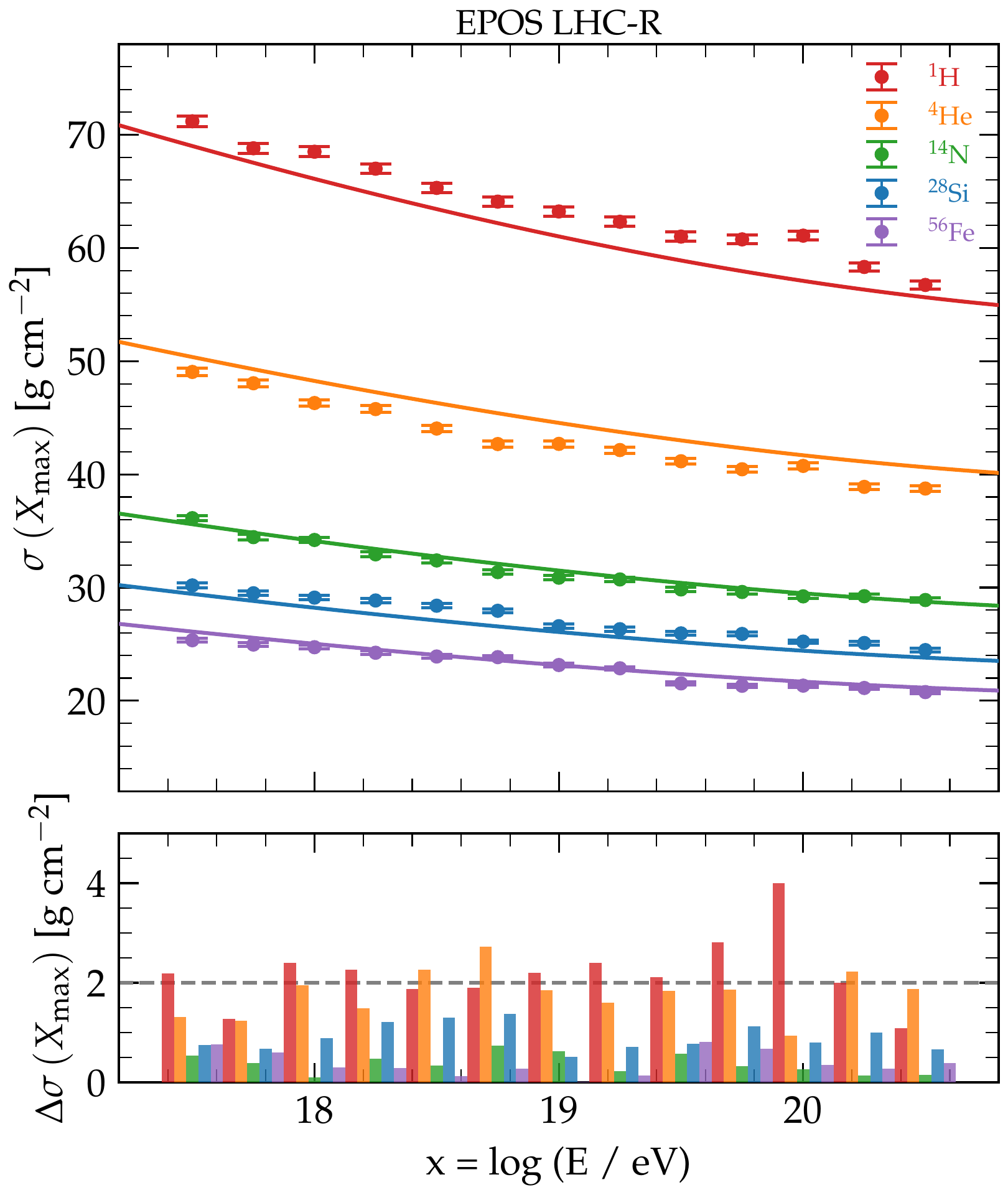}
\includegraphics[width=0.32\textwidth]{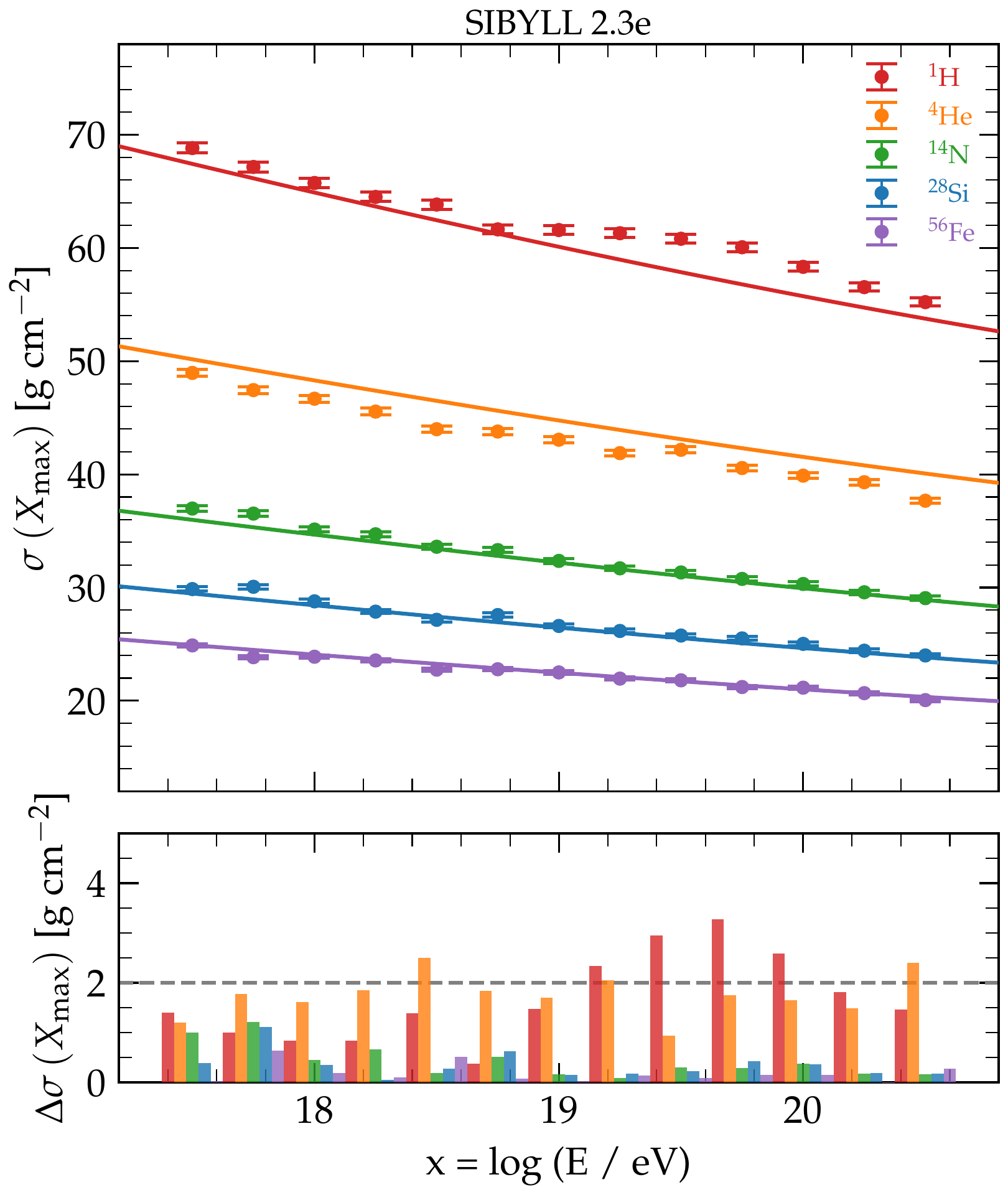}
\includegraphics[width=0.32\textwidth]{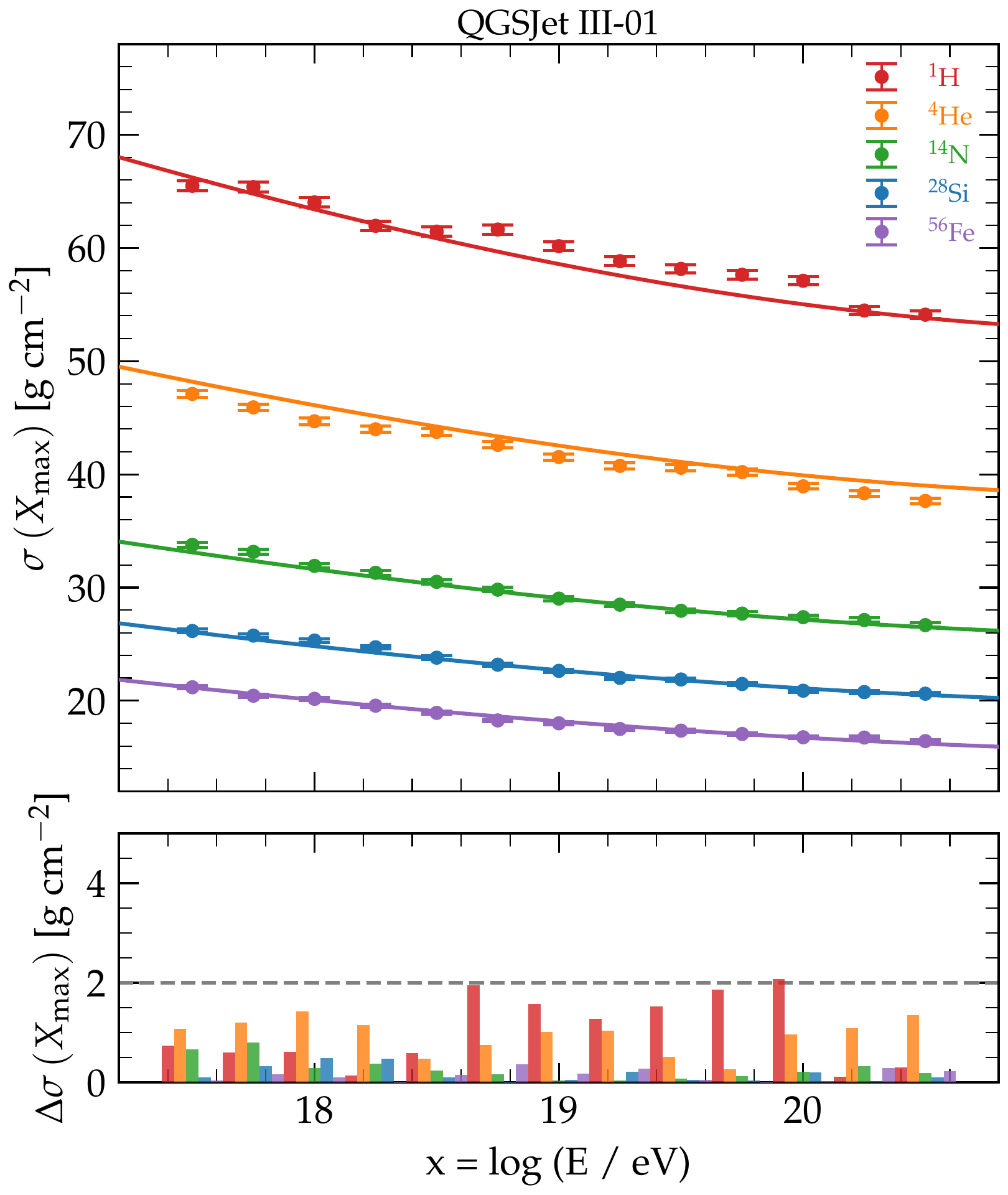}
\caption{Same as Fig.~\ref{fig:xmax_residual}, but for $\sigma(X_{\mathrm{max}})$ using the 
second-order polynomial parametrization of Eq.~\ref{eq:xmax_sigma2}.}
\label{fig:xmax_residual_2nd}
\end{figure}

\begin{figure}[t]
\centering
\includegraphics[width=0.32\textwidth]{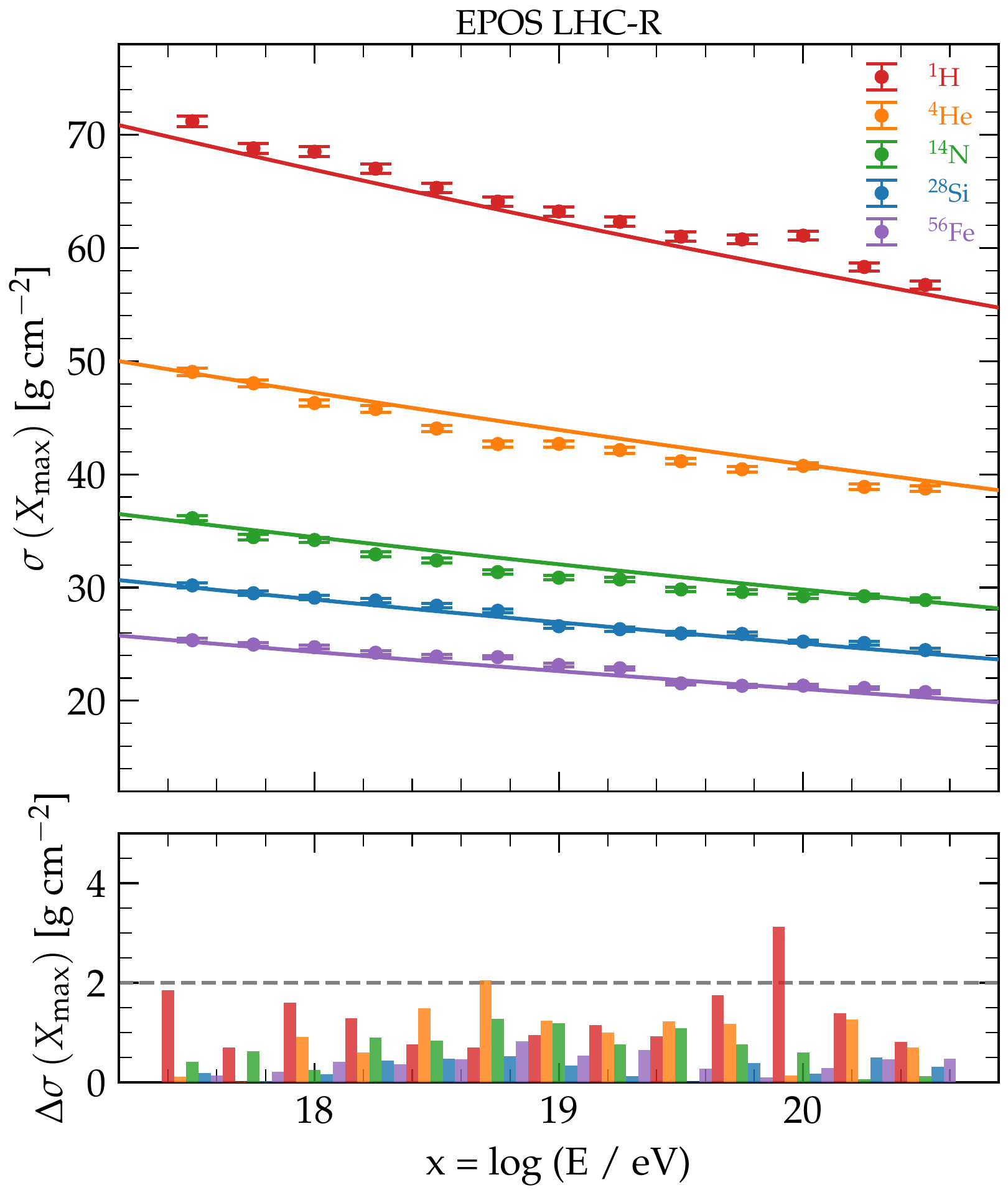}
\includegraphics[width=0.32\textwidth]{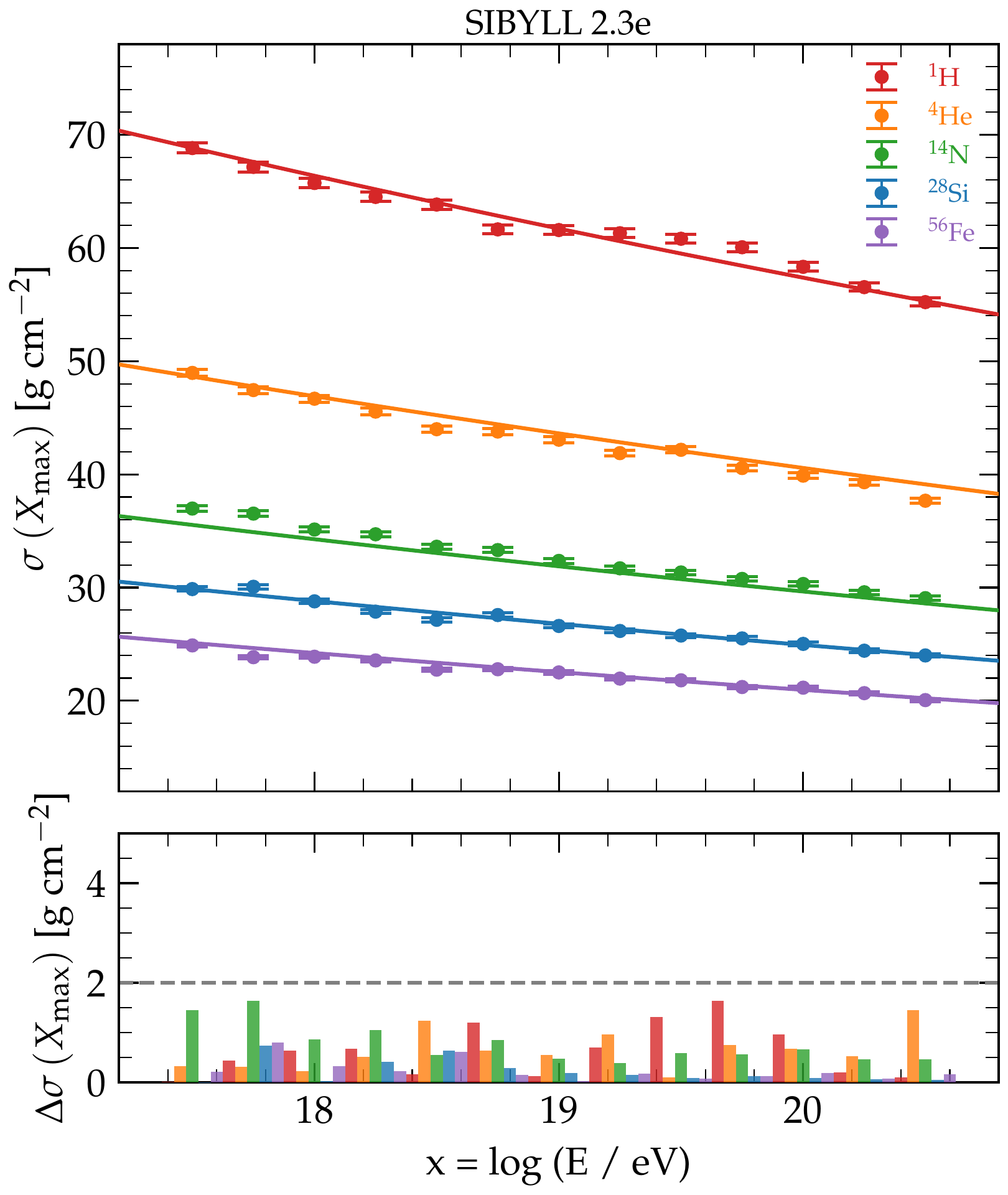}
\includegraphics[width=0.32\textwidth]{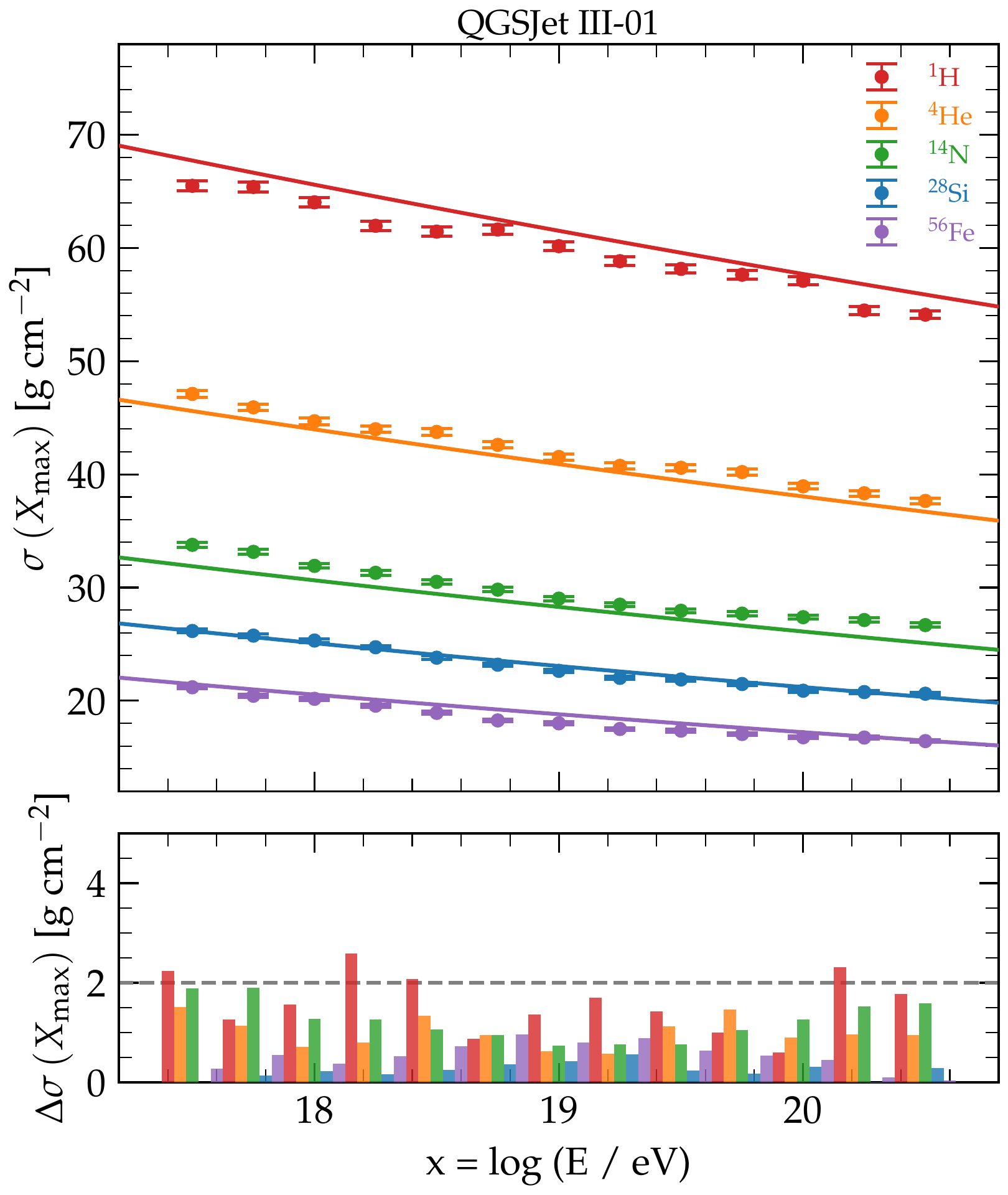}
\caption{Same as Fig.~\ref{fig:xmax_residual}, but for $\sigma(X_{\mathrm{max}})$ using the 
exponential parametrization of Eq.~\ref{eq:xmax_sigma2_exp}.}
\label{fig:xmax_residual_exp}
\end{figure}

To validate the parametrizations, we compare them directly with the results of the \conex~simulations. 
Figure~\ref{fig:xmax_residual} shows, for representative primary nuclei, the simulated values of $\mu_ X$ and $\sigma_X$ together with the corresponding analytical parametrizations. 
In the lower panels of each figure we report the residuals, defined as the absolute difference between the simulation result and the parametrization:
\begin{equation}\label{eq:bias}
\Delta(F) = \left| F_{\mathrm{CONEX}} - F_{\mathrm{parametrization}} \right| .
\end{equation}
This representation gauges the accuracy of the fits across the full energy range. 
In addition to the graphical comparisons, we provide a quantitative summary of the residuals. 
For each hadronic interaction model and primary species, we compute the average residual in the energy range $E \geq 10^{17.5}~\mathrm{eV}$. 
These values are reported in Tables~\ref{tab:xmax_mean_residuals} and~\ref{tab:xmax_sigma_residuals}, for $\langle X_{\mathrm{max}} \rangle$ and $\sigma(X_{\mathrm{max}})$, respectively.

The average residuals for $\langle X_{\mathrm{max}} \rangle$ are found to be small, typically at the level of $1$–$2~\mathrm{g\,cm^{-2}}$, confirming that the parametrizations reproduce the simulation results with high accuracy across the range relevant for composition studies.

The standard deviance, on the other hand, exhibits a systematic bias of about $1$–$2~\mathrm{g\,cm^{-2}}$ for protons and helium across the entire energy range when using the second-order polynomial model in Eq.~\eqref{eq:xmax_sigma2}, corresponding to a relative difference of roughly 3–4\%. 
Employing exponential functions, as in the model of Eq.~\eqref{eq:xmax_sigma2_exp}, reduces the bias to below 2\% for \epos~and \sibyll, while it slightly worsens the bias in the case of \qgsjet. However, we stress that adopting the second-order polynomial functional form preserves the analytical invertibility of the relation, which is required to derive $\ln A$ from the measurement of $\sigma(X_{\mathrm{max}})$, as discussed in Sec.~2 of~\cite{PierreAuger:2013xim} and in Sec.~\ref{sec:lnainversion}. 

\section{Mapping between $X_{\max}$ moments and $\ln A$ moments}
\label{sec:lnainversion}

In composition analyses, it is customary to characterize the $X_{\max}$
 distribution at fixed energy by its first two moments: the mean $\langle X_{\max}\rangle$ and the variance $\sigma^2(X_{\max})$.
When adopting the model-dependent parameterizations introduced in Sec. \ref{sec:xmax}, it is often more convenient to express the composition in terms of the mean logarithmic mass and its variance. The mean logarithmic mass follows directly from Eq. (\ref{eq:xmax_mean}). For greater transparency and ease of interpretation, we rewrite this relation as in~\cite{PierreAuger:2013xim}
\begin{equation}
\label{eq:xmax_linear}
\langle X_{\max}\rangle_Y = X_p(\epsilon) + f_E(\epsilon)\,Y\,,
\end{equation}
where  $X_p = a^\prime(\epsilon)$ denotes the proton $\langle X_{\rm max} \rangle$, and
$f_E = b^\prime(\epsilon)$ is the (energy–dependent) slope relating shifts in $\ln A$ to shifts in $X_{\max}$.
Averaging over the mass distribution yields
\begin{equation}
\label{eq:xmax_mean_lnA}
\langle X_{\max}\rangle
= X_p(\epsilon) + f_E(\epsilon)\,\langle Y\rangle
\quad\implies\quad
\langle Y\rangle
= \frac{\langle X_{\max}\rangle - X_p(\epsilon)}  {f_E(\epsilon)}\,.
\end{equation}

In the case of a single mass having logarithmic value $Y$, the variance of $X_{\max}$ is entirely determined by intrinsic shower-to-shower fluctuations, so that
$$\sigma_X^2(\epsilon,Y) =\sigma_{\rm sh}^2(\epsilon,Y).$$ 
For mixed mass compositions, however, the total variance receives contributions from both intrinsic shower fluctuations and mass mixing. Applying the law of total variance, one obtains \begin{equation}
\label{eq:var_total}
\sigma_X^2 (\epsilon,\langle Y \rangle) 
= \Big\langle \sigma_{\rm sh}^2(\epsilon,Y) \Big\rangle
+ f_E^2(\epsilon)\,\sigma_Y^2\,.
\end{equation}

Equation~\eqref{eq:xmax_mean_lnA} provides a direct mapping between $\langle X_{\max}\rangle$ and $\langle \ln A\rangle$, while Eq.~\eqref{eq:var_total} relates the measured variance of $X_{\max}$ to the mass dispersion $\sigma^2(\ln A)$ once a model for $\sigma_{\rm sh}^2(\epsilon,Y)$ is specified.

\paragraph{Quadratic model for intrinsic fluctuations}

In~\cite{PierreAuger:2013xim} it is proposed to model the intrinsic variance as a quadratic function of $Y$, as given by (\ref{eq:xmax_sigma2})
\begin{equation}
\label{eq:sigma_sh_quad}
\sigma_{\rm sh}^2(\epsilon,Y)
 = \sigma^2_p(\epsilon) \left[ 1 + d^\prime(\epsilon)\, Y + d_2 \, Y^2 \right] ,
\end{equation}
where the parameter $c^\prime$ has been replaced by $\sigma^2_p$ in order to highlight its physical meaning as the intrinsic shower-to-shower fluctuations for protons.

Using $\langle Y^2\rangle = \langle Y \rangle^2 + \sigma_Y^2$, Eq.~\eqref{eq:var_total} becomes
\begin{equation}
\label{eq:sigmaX_quad}
\sigma_X^2
= \sigma^2_p \left[ 1 + d^\prime(\epsilon)\, \langle Y \rangle + d_2 \, \langle Y \rangle^2 \right] 
+ (d_2 \sigma^2_p + f_E^2)\,\sigma_Y^2\,.
\end{equation}
Replacing the first addendum with $\sigma_{\rm sh}^2(\epsilon,\langle Y \rangle)$, which represents the fluctuations associated with a logarithmic mass $\langle Y \rangle)$, one finally gets
\begin{equation}
\label{eq:sigmaY_quad_solution}
\sigma_Y^2
= \frac{\sigma_X^2 - \sigma_{\rm sh}^2(\epsilon,\langle Y \rangle)}{d_2 \sigma^2_p+ f_E^2}\,.
\end{equation}

\paragraph{Exponential model for intrinsic fluctuations}

In the present work we also consider an alternative description of the intrinsic fluctuations, in which the shower variance depends exponentially on $Y$, as in (\ref{eq:xmax_sigma2_exp})
\begin{equation}
\label{eq:sigma_sh_exp}
\sigma^2_{\rm sh}(\epsilon,Y)
= \exp\!\big[f^\prime(\varepsilon) + g^\prime(\epsilon)\,Y\big]\,.
\end{equation}
In this case Eq.~\eqref{eq:var_total} becomes
\begin{equation}
\label{eq:var_total_exp_start}
\sigma_X^2
= \Big\langle \exp\big[f^\prime(\epsilon) + g^\prime(\epsilon)\,Y\big] \Big\rangle
+ f_E^2(\epsilon)\,\sigma_Y^2\,.
\end{equation}

The average of the exponential function depends on the probability distribution function of $Y$, $p(Y)$. In our case, this function is unknown, but its first two moments
$\langle Y \rangle$ and $\sigma_Y^2$ are precisely the quantities we wish to infer. A natural choice is therefore to approximate $p(Y)$ by a Gaussian
\begin{equation}
\label{eq:gaussian_Y}
p(Y) \simeq \mathcal{N}(Y;\,\langle Y \rangle,\sigma_Y^2)\,,
\end{equation}
which is fully determined by those two moments. Under this assumption one obtains
\begin{equation}
\label{eq:exp_average}
\Big\langle \exp\big(f^\prime + g^\prime Y\big)\Big\rangle
= \exp\!\big[f^\prime +  g^\prime \langle Y \rangle)\big]\;\exp\!\Big(\frac{1}{2}{g^\prime}^2\sigma_Y^2\Big)\,,
\end{equation}
where, for compactness, we have omitted the explicit dependence on $\epsilon$.

Inserting Eq.~\eqref{eq:exp_average} into Eq.~\eqref{eq:var_total_exp_start} leads to the implicit equation
\begin{equation}
\label{eq:implicit_sigmaY_exp}
\sigma_X^2
- \sigma^2_{\rm sh}(\epsilon,\langle Y \rangle) 
  \exp\!\Big(\frac{1}{2} {g^\prime}^2\sigma_Y^2\Big)\,
- f_E^2\,\sigma_Y^2 = 0\,,
\end{equation}
to be solved for $\sigma_Y^2$. Unlike the quadratic case, $\sigma_Y^2$ now appears both linearly and in the exponent, so the inversion is no longer a simple algebraic expression. In practice, Eq.~\eqref{eq:implicit_sigmaY_exp} can be solved numerically with standard root–finding algorithms. 

Eq.~\eqref{eq:implicit_sigmaY_exp} can also be solved in closed form using the Lambert $W$ function as
\begin{equation}
\label{eq:sigmaY_exp_Lambert}
\sigma_Y^2
= \frac{\sigma_X^2}{f_E^2}
- \frac{2}{{g^\prime}^2}\,
W_0\!\left(
\frac{{g^\prime}^2}{2f_E^2}\,
\sigma^2_{\rm sh}(\epsilon,\langle Y \rangle)  \exp \left[ \frac{{g^\prime}^2}{2 f_E^2}\,\sigma_X^2
\right]
\right),
\end{equation}
where $W_0$ is the principal branch of the Lambert $W$ function.

Given these ingredients, the exponential model provides a consistent forward description of
$\sigma_{\rm sh}^2(\epsilon,Y)$ and, via Eq.~\eqref{eq:implicit_sigmaY_exp} (or Eq.~\eqref{eq:sigmaY_exp_Lambert}), an in–principle invertible mapping between the measured $X_{\max}$ moments and the underlying $\ln A$ moments.

\section{Parameterization of $X_{\mathrm{max}}$ distributions}
\label{sec:gumbel}

Beyond the first two moments, a full description of the $X_{\mathrm{max}}$ distributions is required for composition analyses that exploit the full shape of the profiles. Following the approach introduced in Ref.~\cite{DeDomenico:2013wwa}, we model the simulated $X_{\mathrm{max}}$ distributions with generalized Gumbel functions. This method has been extensively used in combined fits of the energy spectrum and mass composition of UHECR measurements (see e.g.~\cite{PierreAuger:2016use,PierreAuger:2024hlp}) and provides a compact analytical representation of the distributions for all hadronic interaction models considered in this work. 

The functional form of the generalized Gumbel distribution is
\begin{equation}
\mathcal G\!\left(X; \mu,\sigma,\lambda\right) 
= \frac{1}{\sigma}\,\frac{\lambda^{\lambda}}{\Gamma(\lambda)}
\exp\!\left[-\lambda\left(z + e^{-z}\right)\right],
\qquad 
z \equiv \frac{X - \mu}{\sigma}, 
\label{eq:gumbel}
\end{equation}
with $\sigma>0$ and $\lambda>0$. Here $\mu$ is the location parameter, $\sigma$ the scale, and $\lambda$ a shape parameter controlling the skewness of the distribution. For $\lambda=1$, Eq.~\eqref{eq:gumbel} reduces to the standard (type-I) Gumbel distribution. 
This parametrization has been shown to perform better than other proposed functional forms in reproducing both the asymmetric shape and the tail behaviour of the $X_{\max}$ distributions~\cite{Arbeletche:2019qma}.

Each of the Gumbel parameters is expressed as a smooth function of the logarithmic energy  
\mbox{$\epsilon \equiv \log_{10}(E/E_0)$}, with $E_0 = 10^{19}$~eV, and of the logarithm of the mass number 
\mbox{$Y \equiv \ln A$}. In particular, the location parameter $\mu$ is modeled as a quadratic function of $\epsilon$, with coefficients that are themselves quadratic functions of $Y$. The scale $\sigma$ and shape $\lambda$ parameters are modeled as linear functions of $\epsilon$, again with $Y$-dependent quadratic coefficients. Explicitly:
\begin{itemize} 

\item Location $\mu(\epsilon,Y)$ (quadratic in $\epsilon$):
\begin{eqnarray}
p^\mu_0(Y) &=& a^\mu_0 + a^\mu_1 \, Y + a^\mu_2 \, Y^2, \nonumber\\
p^\mu_1(Y) &=& b^\mu_0 + b^\mu_1 \, Y + b^\mu_2 \, Y^2, \nonumber \\
p^\mu_2(Y) &=& c^\mu_0 + c^\mu_1 \, Y + c^\mu_2 \, Y^2, \nonumber \\
\mu(\epsilon,Y) &=& p^\mu_0(Y) + p^\mu_1(Y)\,\epsilon + p^\mu_2(Y)\,\epsilon^2 \, .
\label{eq:mu}
\end{eqnarray}

\item Scale $\sigma(\epsilon,Y)$ (linear in $\epsilon$):
\begin{eqnarray}
p^\sigma_0(Y) &=& a^\sigma_0 + a^\sigma_1 \, Y + a^\sigma_2 \, Y^2, \nonumber \\
p^\sigma_1(Y) &=& b^\sigma_0 + b^\sigma_1 \, Y + b^\sigma_2 \, Y^2, \nonumber \\
\sigma(\epsilon,Y) &=& p^\sigma_0(Y) + p^\sigma_1(Y)\,\epsilon \, .
\label{eq:sigma}
\end{eqnarray}

\item Shape $\lambda(\epsilon,Y)$ (linear in $\epsilon$):
\begin{eqnarray}
p^\lambda_0(Y) &=& a^\lambda_0 + a^\lambda_1 \, Y + a^\lambda_2 \, Y^2, \nonumber \\
p^\lambda_1(Y) &=& b^\lambda_0 + b^\lambda_1 \, Y + b^\lambda_2 \, Y^2, \nonumber \\
\lambda(\epsilon,Y) &=& p^\lambda_0(Y) + p^\lambda_1(Y)\,\epsilon \, .
\label{eq:lambda}
\end{eqnarray}
\end{itemize}
In total, this scheme involves 9 coefficients for $\mu$, and 6 coefficients each for $\sigma$ and $\lambda$, yielding $21$ free parameters.

To determine these coefficients, we perform a global unbinned maximum-likelihood fit to the full set of simulated $X_{\mathrm{max}}$ values for each hadronic interaction model. Each simulated shower is characterized by its energy $E_i$, mass number $A_i$ and depth of shower maximum $X_{\mathrm{max},i}$, from which we construct
\( \epsilon_i \equiv \log_{10}(E_i/E_0) \), and \( Y_i \equiv \ln A_i \).

For given values of the parameter vector $\boldsymbol{\theta}$, containing the 21 polynomial coefficients entering Eqs.~\eqref{eq:mu}–\eqref{eq:lambda}, we compute the corresponding Gumbel parameters
$\mu_i \equiv \mu(\epsilon_i,Y_i;\boldsymbol{\theta})$, 
$\sigma_i \equiv \sigma(\epsilon_i,Y_i;\boldsymbol{\theta})$ and 
$\lambda_i \equiv \lambda(\epsilon_i,Y_i;\boldsymbol{\theta})$
for each event $i$. Assuming independent showers, the likelihood for the full data set $\{X_{\mathrm{max},i}\}_{i=1}^N$ is
\begin{equation}
\mathcal{L}(\boldsymbol{\theta})
= \prod_{i=1}^{N}
\mathcal G\!\left[
X_{\mathrm{max},i};\,
\mu_i(\epsilon_i, Y_i; \boldsymbol{\theta}),
\sigma_i(\epsilon_i, Y_i; \boldsymbol{\theta}),
\lambda_i(\epsilon_i, Y_i; \boldsymbol{\theta})
\right],
\label{eq:likelihood}
\end{equation}
and we maximize the corresponding log-likelihood (for compactness, we omit the explicit dependence of $\mu_i,\sigma_i,\lambda_i$ on $\epsilon_i, Y_i, \boldsymbol{\theta}$)
\begin{equation}
\ln \mathcal{L}(\boldsymbol{\theta})
= \sum_{i=1}^{N}
\ln \mathcal G\!\left(
X_{\mathrm{max},i};\,
\mu_i,
\sigma_i,
\lambda_i
\right).
\label{eq:loglikelihood}
\end{equation}

Using the explicit form of the generalized Gumbel density in Eq.~\eqref{eq:gumbel}, the log-likelihood contribution of a single event $i$ can be written as
\begin{equation}
\ln \mathcal G_i
= -\ln \sigma_i
+ \lambda_i \ln \lambda_i
- \ln \Gamma(\lambda_i)
- \lambda_i z_i
- \lambda_i e^{-z_i},
\qquad
z_i \equiv \frac{X_{\mathrm{max},i} - \mu_i}{\sigma_i}.
\end{equation}
It is numerically convenient to minimize the negative log-likelihood 
$\mathrm{NLL}(\boldsymbol{\theta}) \equiv -\ln \mathcal{L}(\boldsymbol{\theta})$, which becomes
\begin{equation}
\mathrm{NLL}(\boldsymbol{\theta})
= \sum_{i=1}^{N}
\Big[
\ln \sigma_i
- \lambda_i \ln \lambda_i
+ \ln \Gamma(\lambda_i)
+ \lambda_i z_i
+ \lambda_i e^{-z_i}
\Big].
\label{eq:nll}
\end{equation}
In the numerical implementation, $\mu_i$, $\sigma_i$ and $\lambda_i$ are computed event by event from the polynomial parameterizations in $\epsilon_i$ and $Y_i$, and the sum in Eq.~\eqref{eq:nll} is evaluated over all simulated showers. 
In this way, all available simulated events contribute directly to the determination of $\boldsymbol{\theta}$ without any loss of information from binning in $X_{\mathrm{max}}$.

\subsection{Fit parameters}

The coefficients entering Eqs.~\eqref{eq:mu}–\eqref{eq:lambda} are obtained by minimizing the negative log-likelihood in Eq.~\eqref{eq:nll} for each hadronic interaction model. The minimization is performed with the \texttt{iminuit}~\cite{iminuit} package, using the default \textsc{Migrad} algorithm, followed by a \textsc{Hesse} step to estimate the covariance matrix and the corresponding parameter uncertainties.

To ensure that the fit remains in a physically meaningful region of parameter space, we impose constraints on the Gumbel parameters implied by a given choice of $\boldsymbol{\theta}$. In particular, we require the scale and shape parameters to satisfy $\sigma_i > 0$ and $\lambda_i > 0$ for all events, and we restrict the dynamically computed values of
$\mu_i$, $\sigma_i$ and $\lambda_i$ to lie within broad but physically motivated intervals, 
$300 < \mu_i < 1300$, $10 < \sigma_i < 100$, and $0.01 < \lambda_i < 10$. Trial parameter sets that violate any of these conditions are assigned a very large value of $\mathrm{NLL}$, so that they are effectively excluded from the region explored by the minimizer. Since Eq.~\eqref{eq:gumbel} defines a properly normalized probability density, no additional free normalization parameter is introduced in the likelihood.

\begin{figure}[t]
\centering
\includegraphics[width=0.32\textwidth]{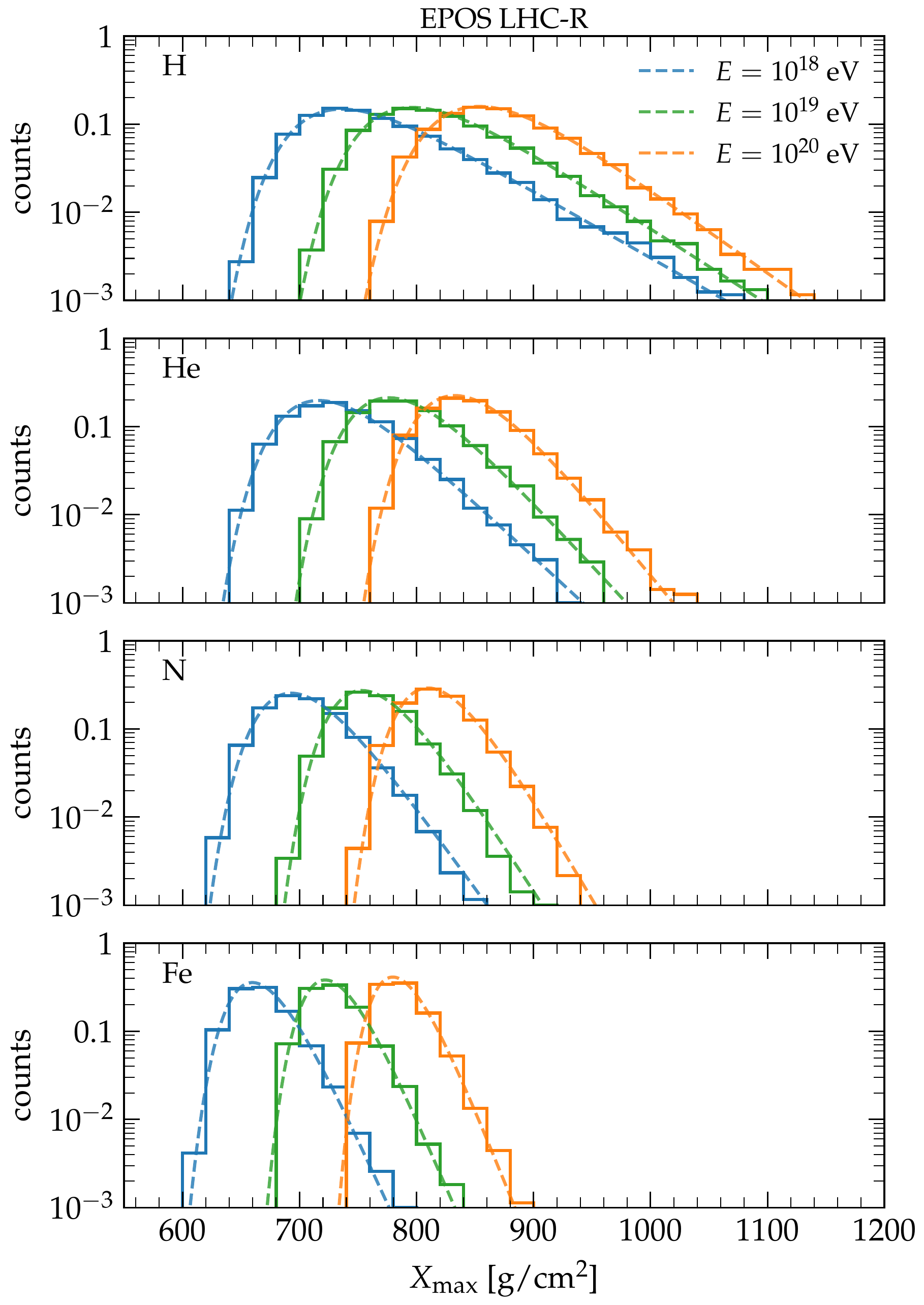}
\includegraphics[width=0.32\textwidth]{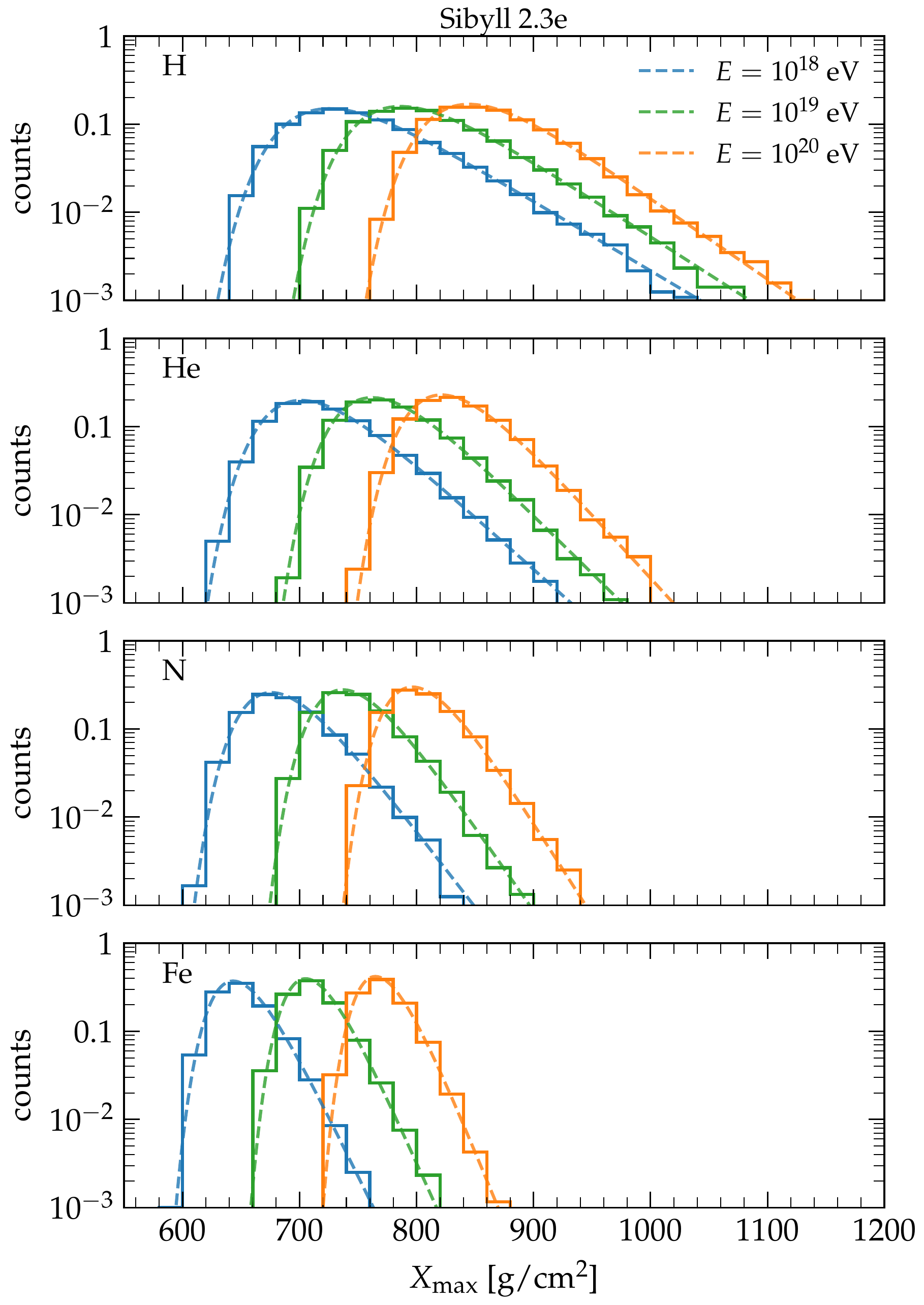}
\includegraphics[width=0.32\textwidth]{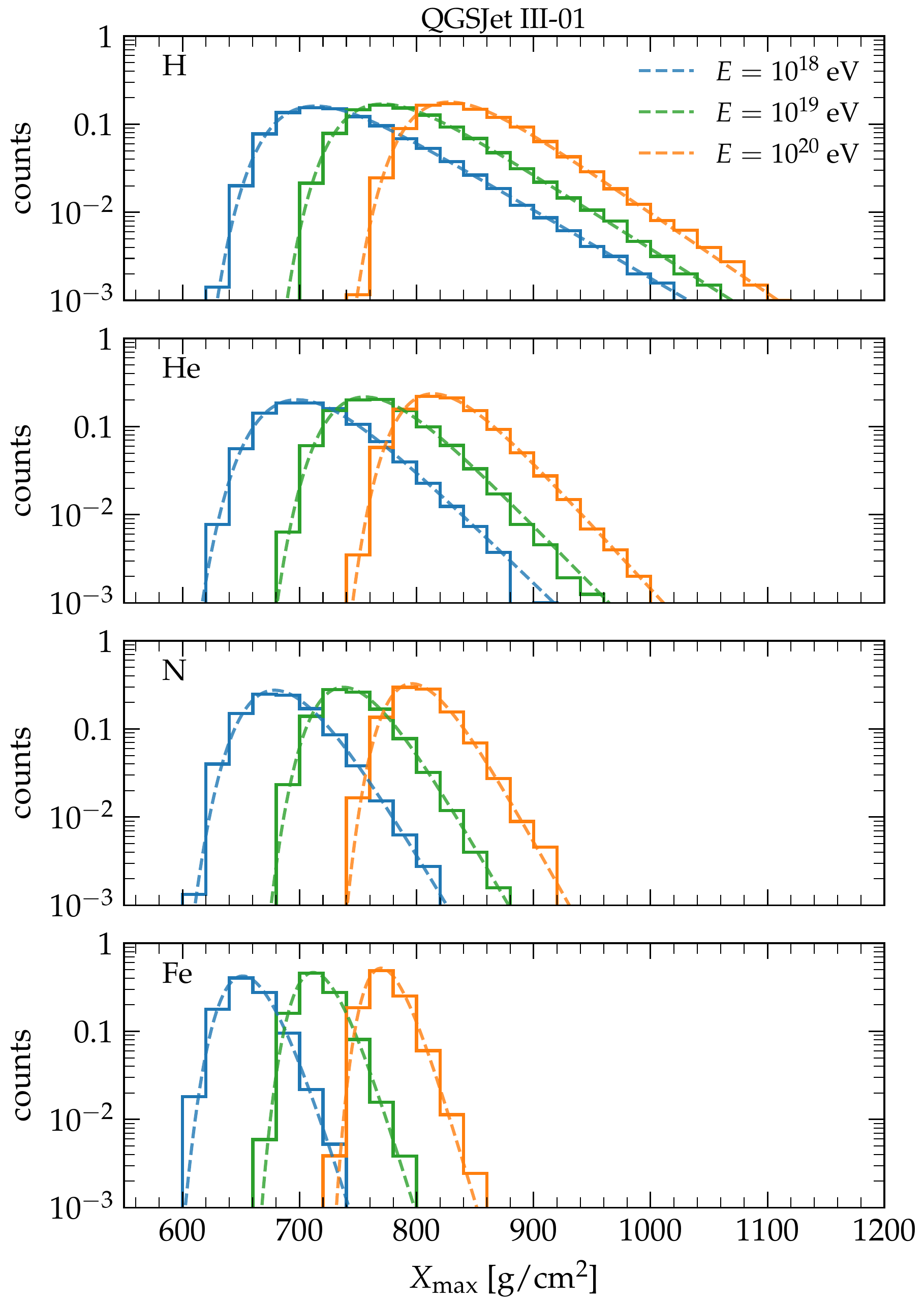}
\caption{Representative examples of simulated $X_{\mathrm{max}}$ distributions from \conex\ (histograms) and the corresponding generalized Gumbel descriptions (solid curves) obtained from the global parameterization of Eqs.~\eqref{eq:mu}--\eqref{eq:lambda}. From left to right, the panels refer to the hadronic interaction models \epos, \sibyll, and \qgsjet. Each panel shows selected primaries (H, He, N, Fe) and energies ($10^{18}$ eV, $10^{19}$ eV, $10^{20}$ eV) to illustrate the quality of the description across the simulated phase space.}
\label{fig:gumbel_fits}
\end{figure}

To mitigate possible dependence on the starting point of the minimization and reduce the risk of converging to local minima, we repeat the fit $N_{\mathrm{seed}} = 100$ times with randomly chosen initial values of $\boldsymbol{\theta}$, sampled within wide ranges around physically reasonable expectations. For each seed, the minimization is iterated until \textsc{Migrad} converges and the \textsc{Hesse} matrix is successfully computed. Among all successful fits, we select as our best solution the one with the smallest deviance,
\begin{equation}
D(\boldsymbol{\theta}) \equiv -2 \ln \mathcal{L}(\boldsymbol{\theta}) \, ,
\end{equation}
which is equivalent to choosing the minimum of the negative log-likelihood up to an irrelevant additive constant.

The best-fit coefficients for the three hadronic interaction models (\sibyll, \epos, \qgsjet) are reported in Table~\ref{tab:gumbel_parameter_fit}. Representative examples of one-dimensional $X_{\mathrm{max}}$ distributions for different primaries and energies are shown in Fig.~\ref{fig:gumbel_fits}, together with the corresponding generalized Gumbel functions evaluated using the global parameterization.

\begin{figure}[t]
\centering
\includegraphics[width=0.32\textwidth]{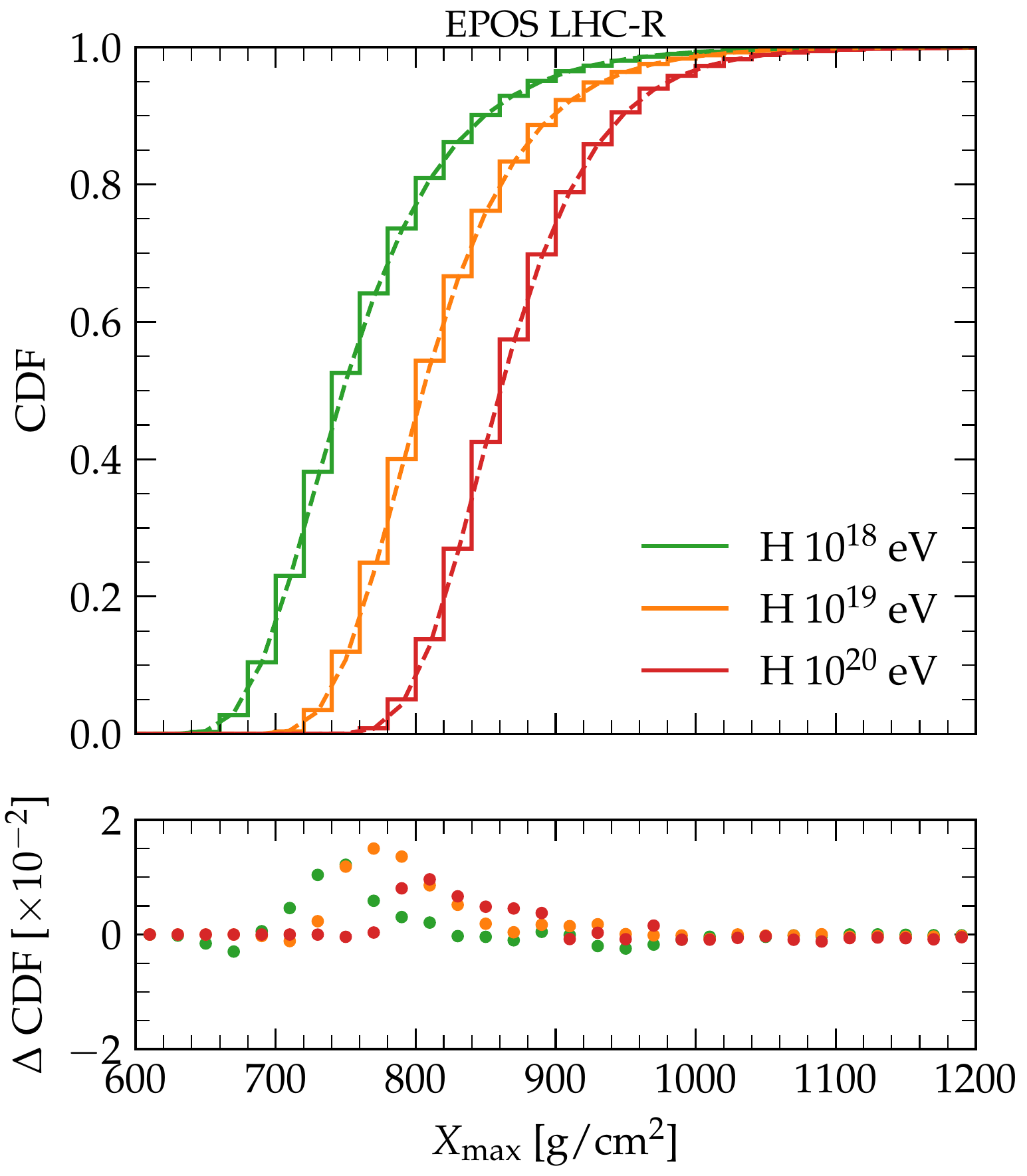}
\includegraphics[width=0.32\textwidth]{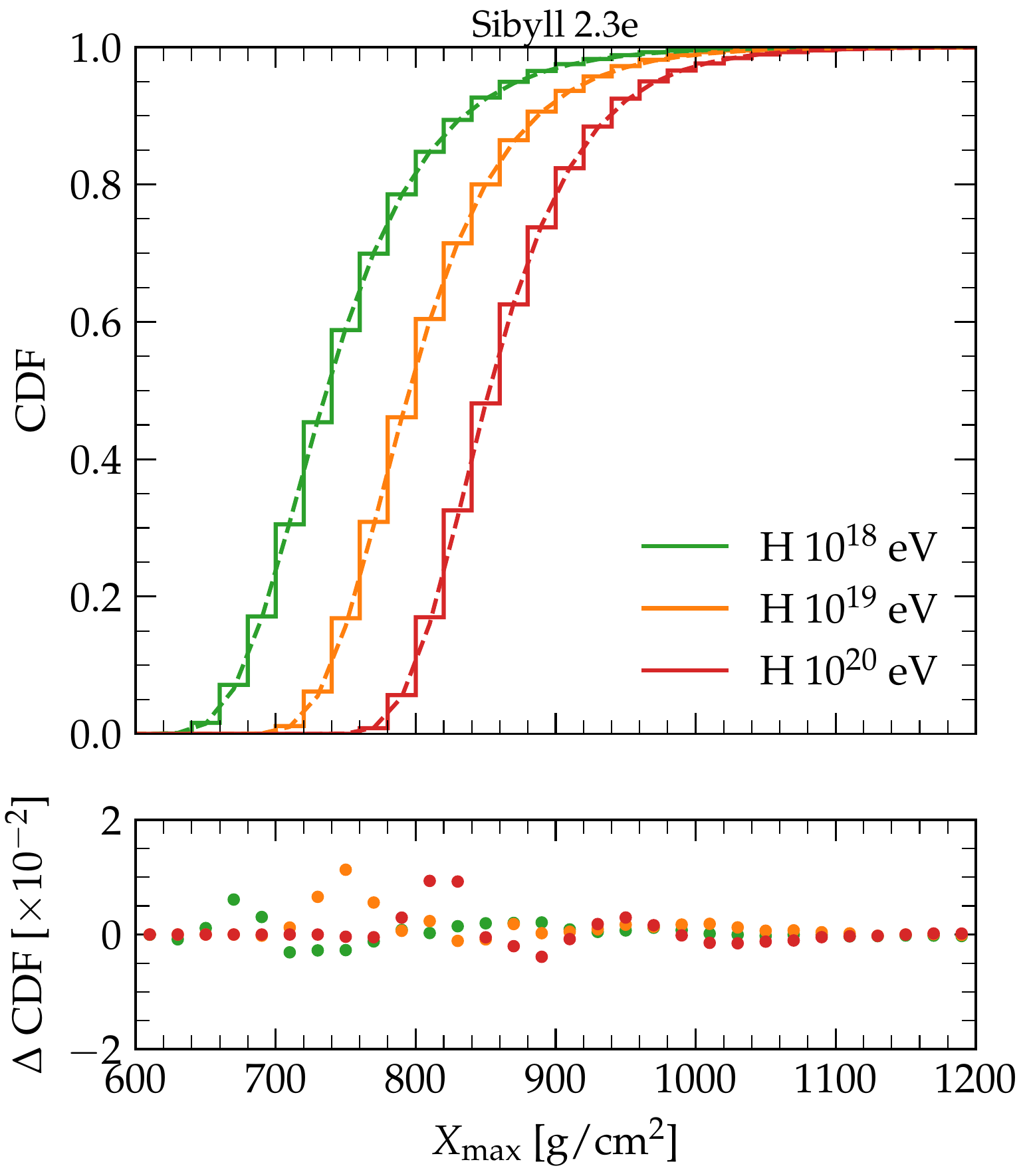}
\includegraphics[width=0.32\textwidth]{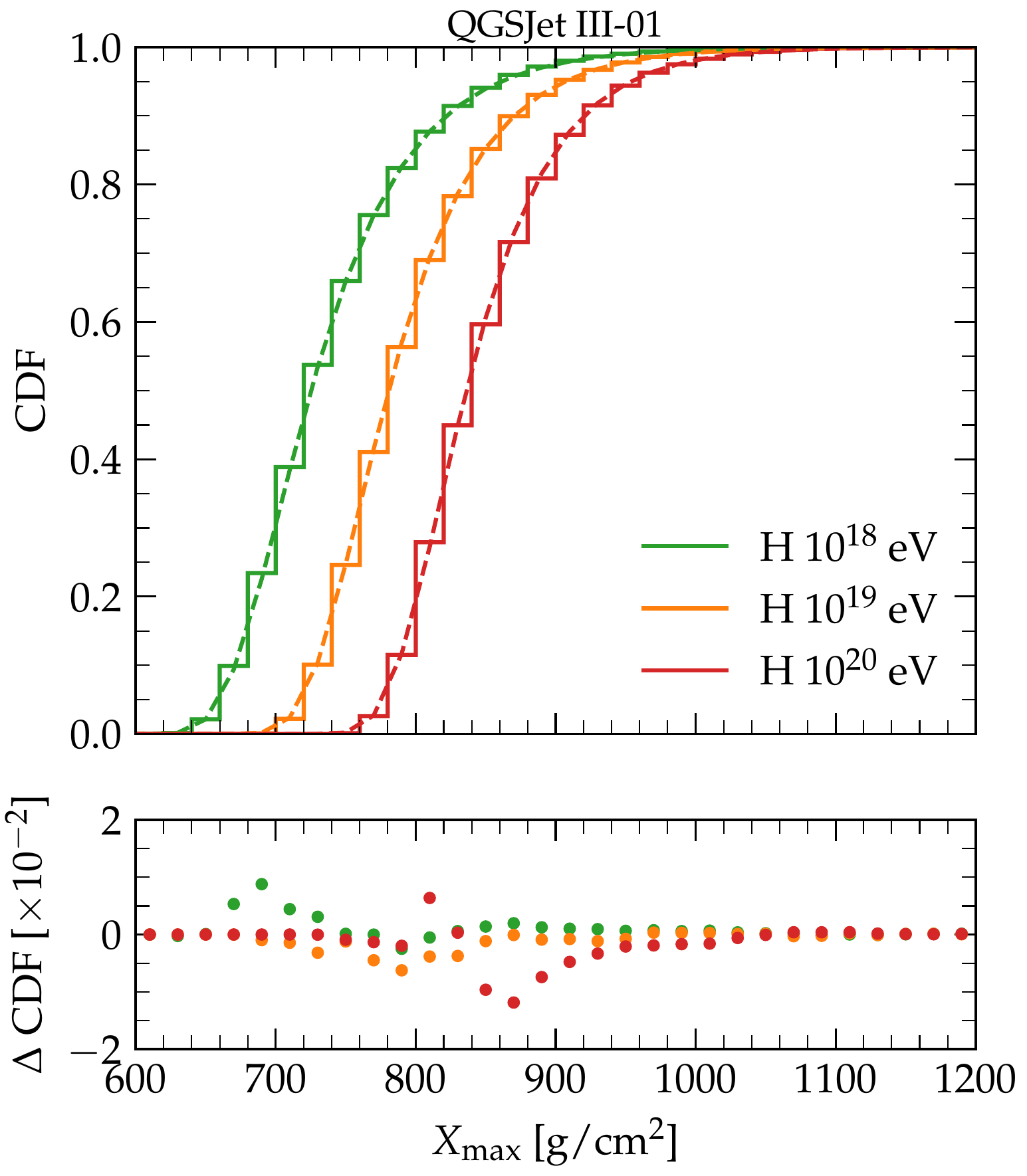}

\includegraphics[width=0.32\textwidth]{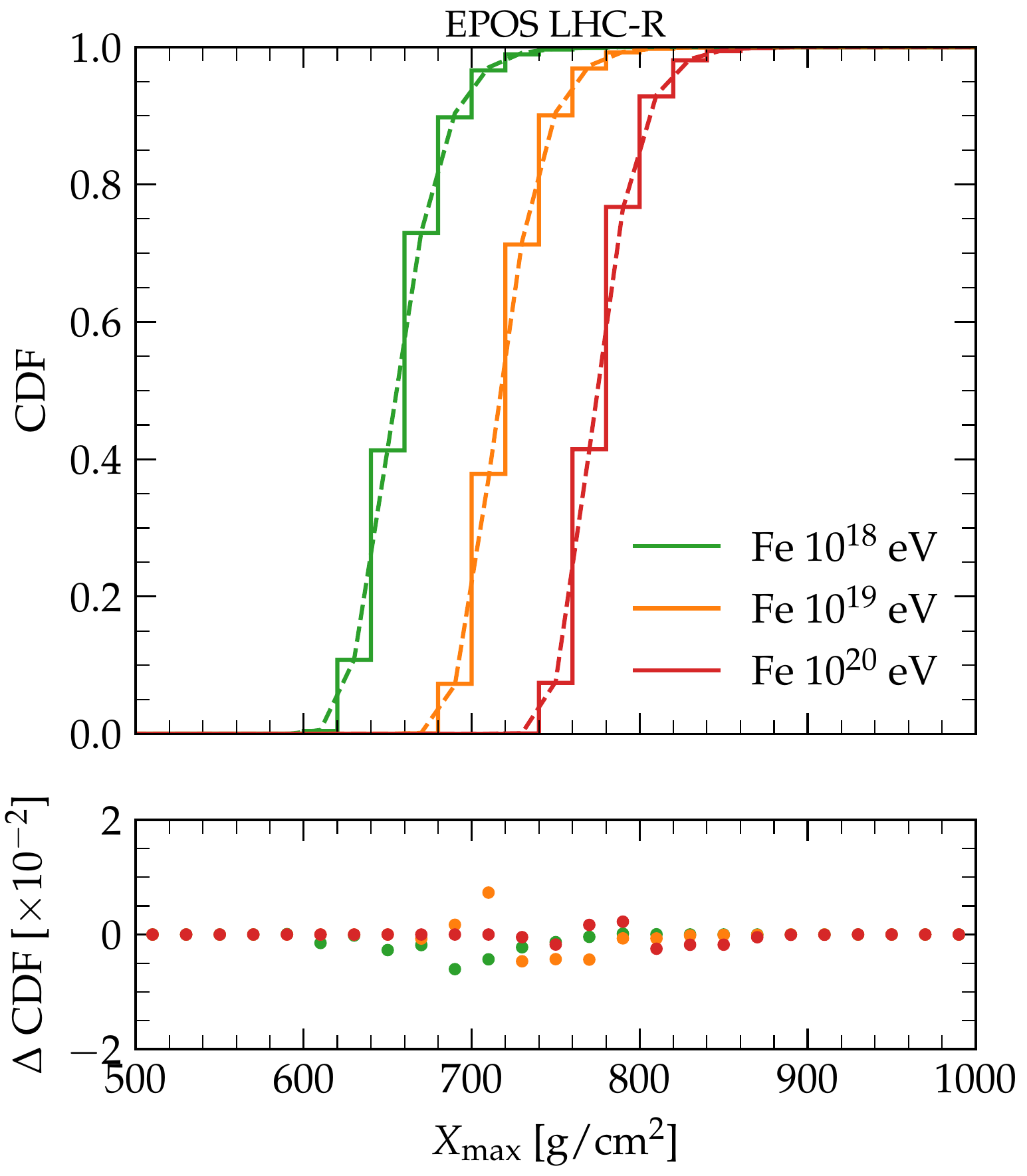}
\includegraphics[width=0.32\textwidth]{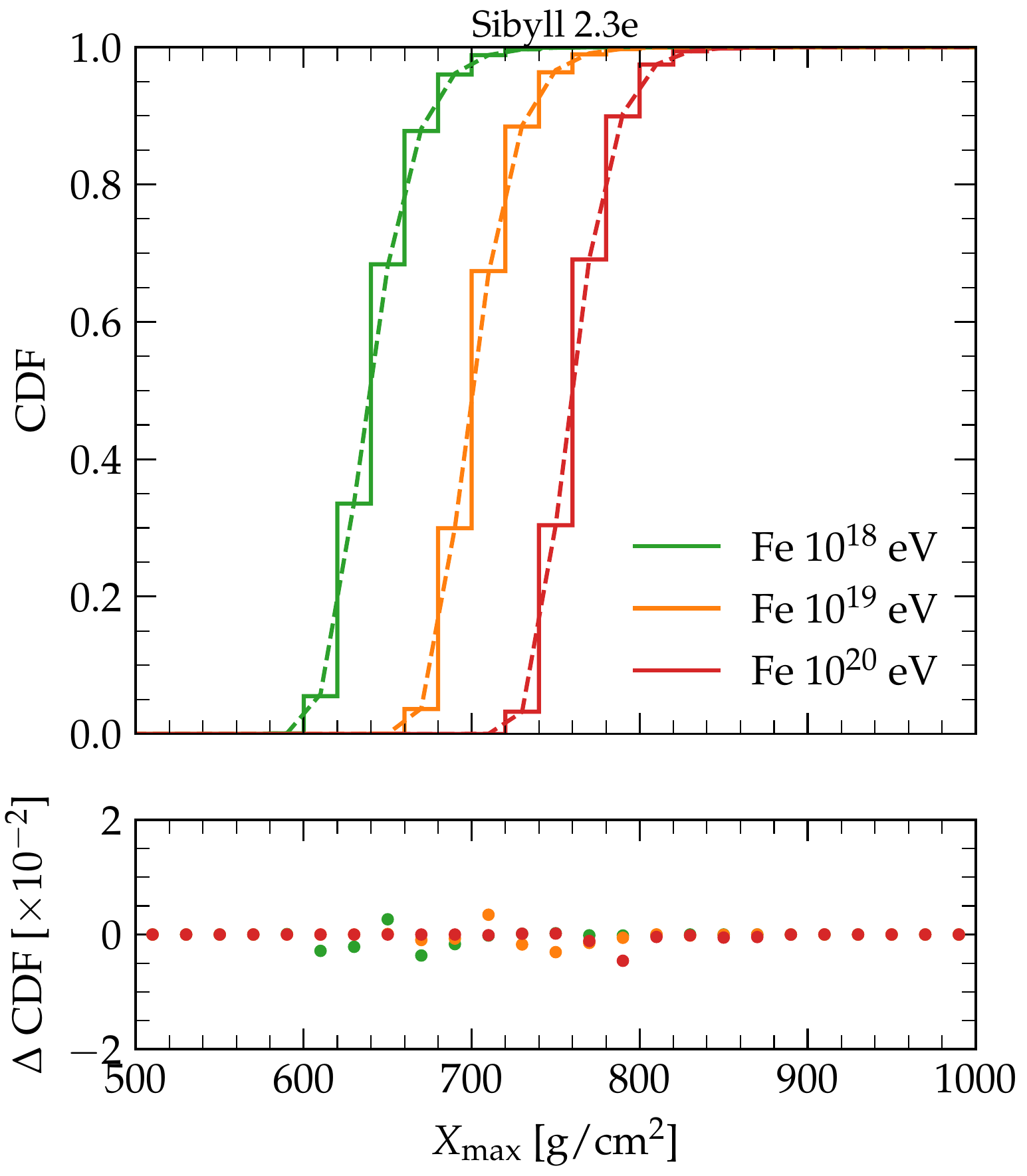}
\includegraphics[width=0.32\textwidth]{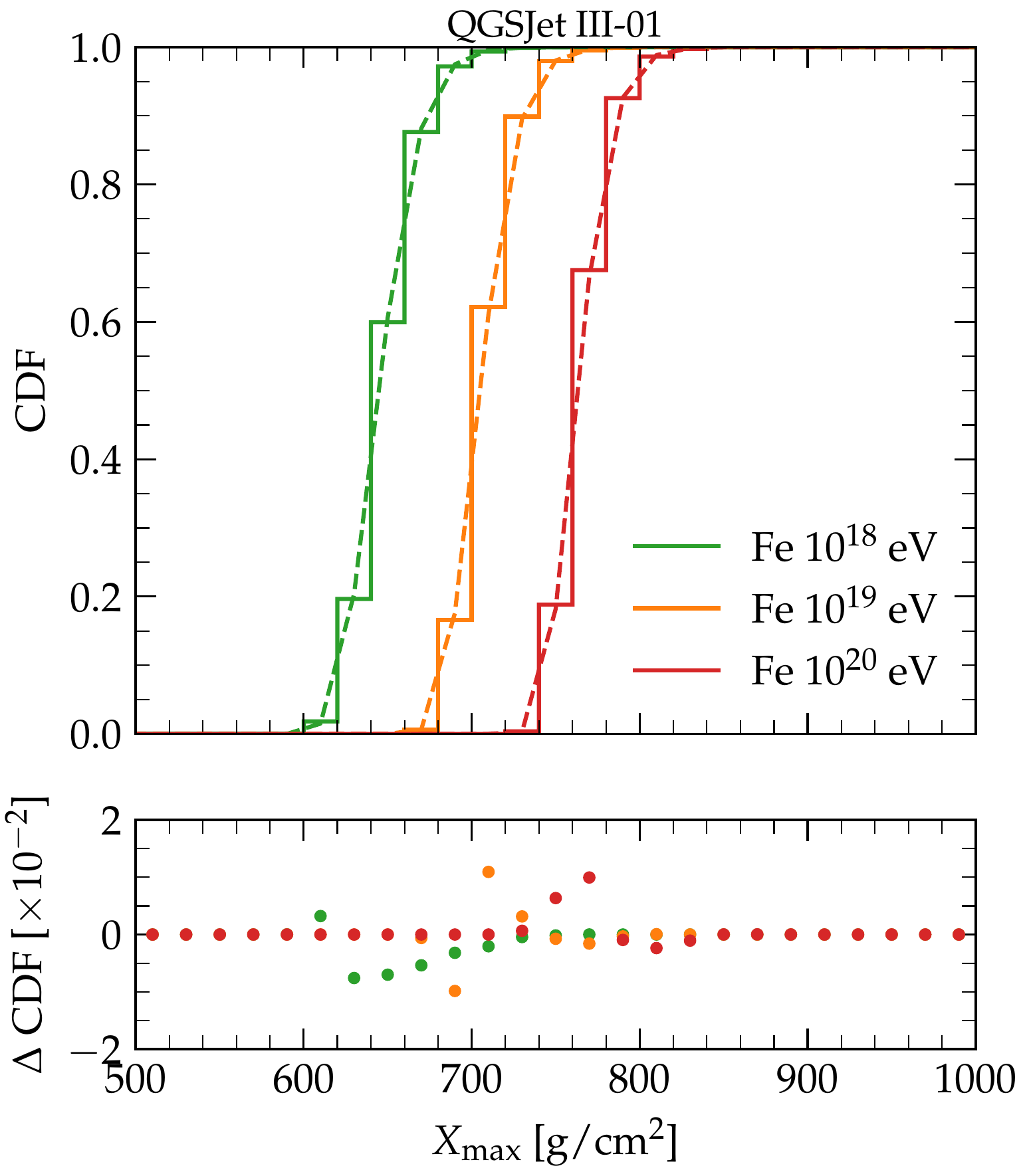}
\caption{Empirical cumulative distribution functions (CDFs) of $X_{\mathrm{max}}$ from \conex\ simulations (histograms) compared to the CDFs predicted by the generalized Gumbel parameterization (solid lines) for H (top row) and Fe (bottom row). In each panel the three energies $E=10^{18}$~eV, $E=10^{19}$~eV and $E=10^{20}$~eV are shown. Columns correspond, from left to right, to the hadronic interaction models \epos, \sibyll, and \qgsjet. The lower sub-panels display the residuals $\Delta\mathrm{CDF}(X)\equiv \mathrm{CDF}_{\mathrm{MC}}(X)-\mathrm{CDF}_{\mathrm{G}}(X)$.}
\label{fig:gumbel_cdf}
\end{figure}

As an additional validation, we compare the cumulative distribution functions (CDFs) obtained from the \conex~simulations with those predicted by the Gumbel parameterization. Specifically, for H and Fe primaries at $E = 10^{18}$~eV and $E = 10^{20}$~eV, Fig.~\ref{fig:gumbel_cdf} displays the empirical CDFs, constructed directly from the ordered $X_{\mathrm{max}}$ values, together with the theoretical CDFs obtained by integrating Eq.~\eqref{eq:gumbel} from $0$ to $X$. The lower panels show the residuals, defined as the difference between the simulated and fitted CDFs. These residuals remain at the few-percent level across the full range of $X_{\mathrm{max}}$, demonstrating that the generalized Gumbel parameterization provides a robust analytical representation of the $X_{\mathrm{max}}$ distributions and is sufficiently accurate for use in combined spectrum and composition analyses.

\section{Consistency of Gumbel representation}
\label{sec:gumbelconsistency}

As an internal consistency check of the generalized Gumbel parameterization, we verify that the first two moments implied by the fitted Gumbel parameters reproduce the mean and variance of the underlying \conex\ samples. For each $(E,A)$ bin, the generalized Gumbel distribution in Eq.~\eqref{eq:gumbel} has an analytical mean and variance given by
\begin{align}
\langle X_{\mathrm{max}} \rangle_{\mathcal{G}} 
&= \mu + \sigma \bigl[\ln \lambda - \psi(\lambda)\bigr] \, , 
\label{eq:gumbel_mean} \\
\mathrm{Var}(X_{\mathrm{max}})_{\mathcal{G}}
&= \sigma^2 \, \psi_1(\lambda) \, , 
\label{eq:gumbel_var}
\end{align}
where $\psi(\lambda)$ is the digamma function and $\psi_1(\lambda)$ is the trigamma function, i.e.\ the first derivative of $\psi(\lambda)$ with respect to $\lambda$.
In our implementation, Eqs.~\eqref{eq:gumbel_mean}–\eqref{eq:gumbel_var} are evaluated using the fitted values of $\mu(\epsilon,Y)$, $\sigma(\epsilon,Y)$ and $\lambda(\epsilon,Y)$ from Eqs.~\eqref{eq:mu}–\eqref{eq:lambda} at the corresponding 
$\epsilon = \log_{10}(E/E_0)$ and $Y = \ln A$.

\begin{figure}[t]
\centering
\includegraphics[width=0.32\textwidth]{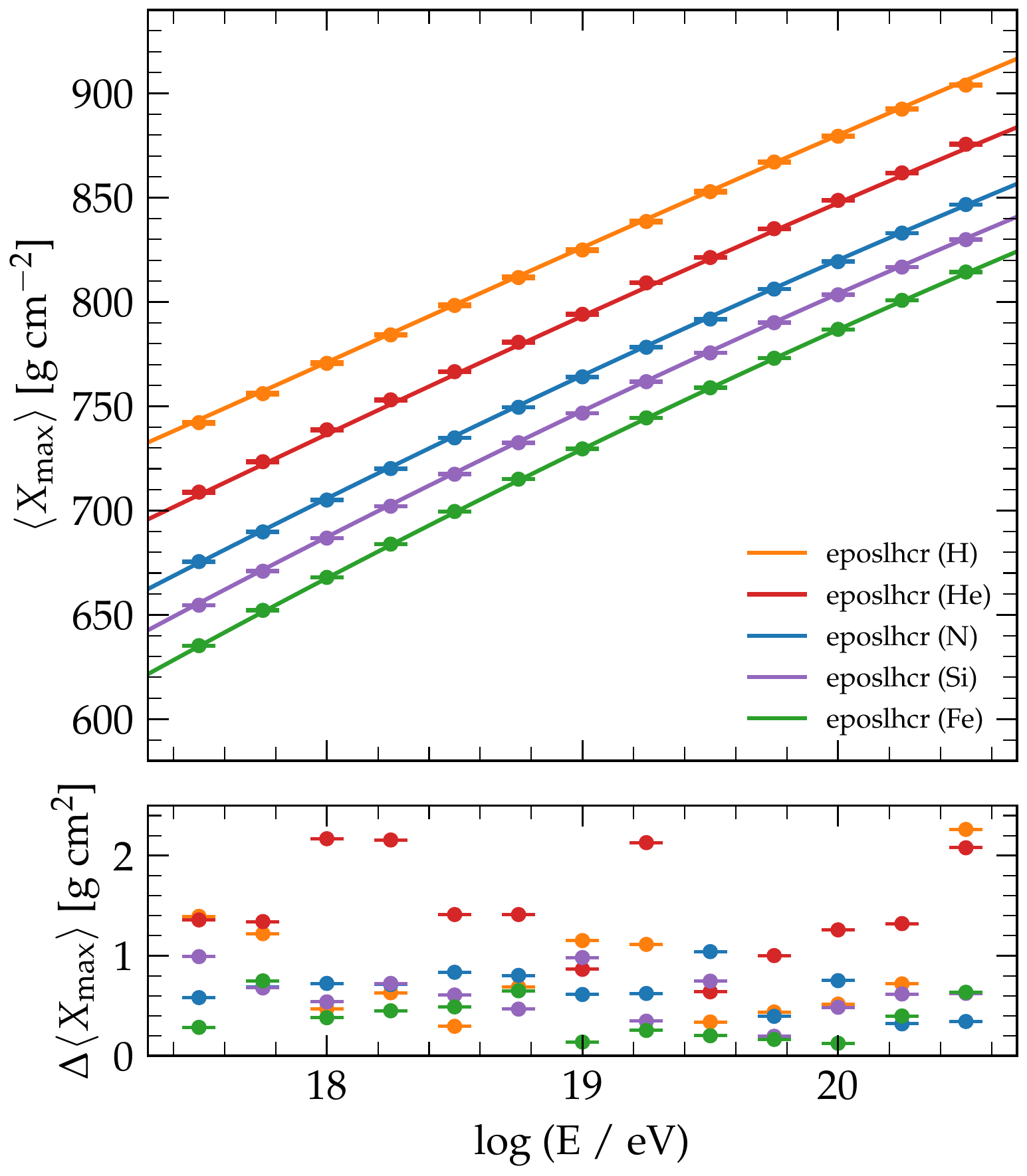}
\includegraphics[width=0.32\textwidth]{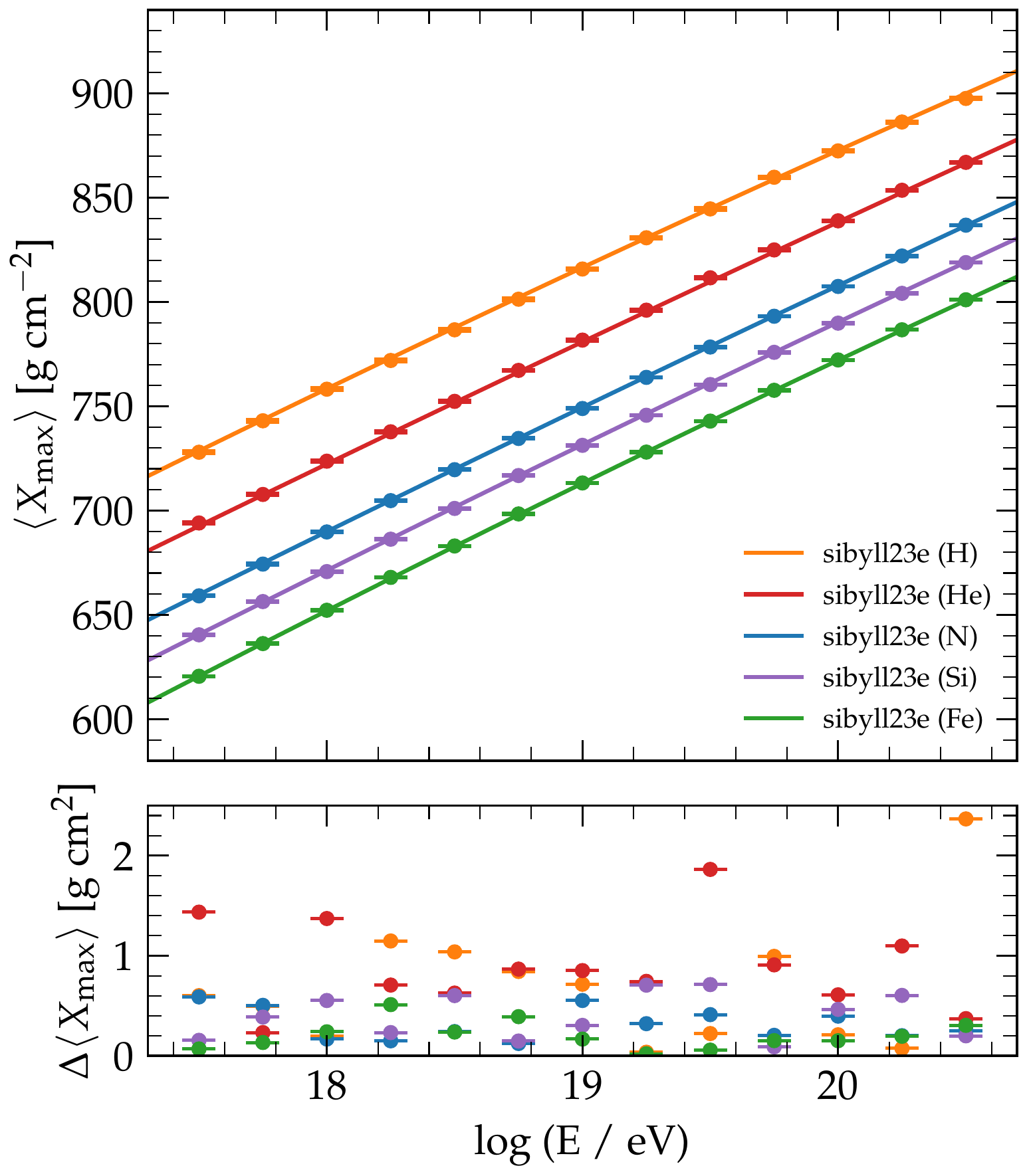}
\includegraphics[width=0.32\textwidth]{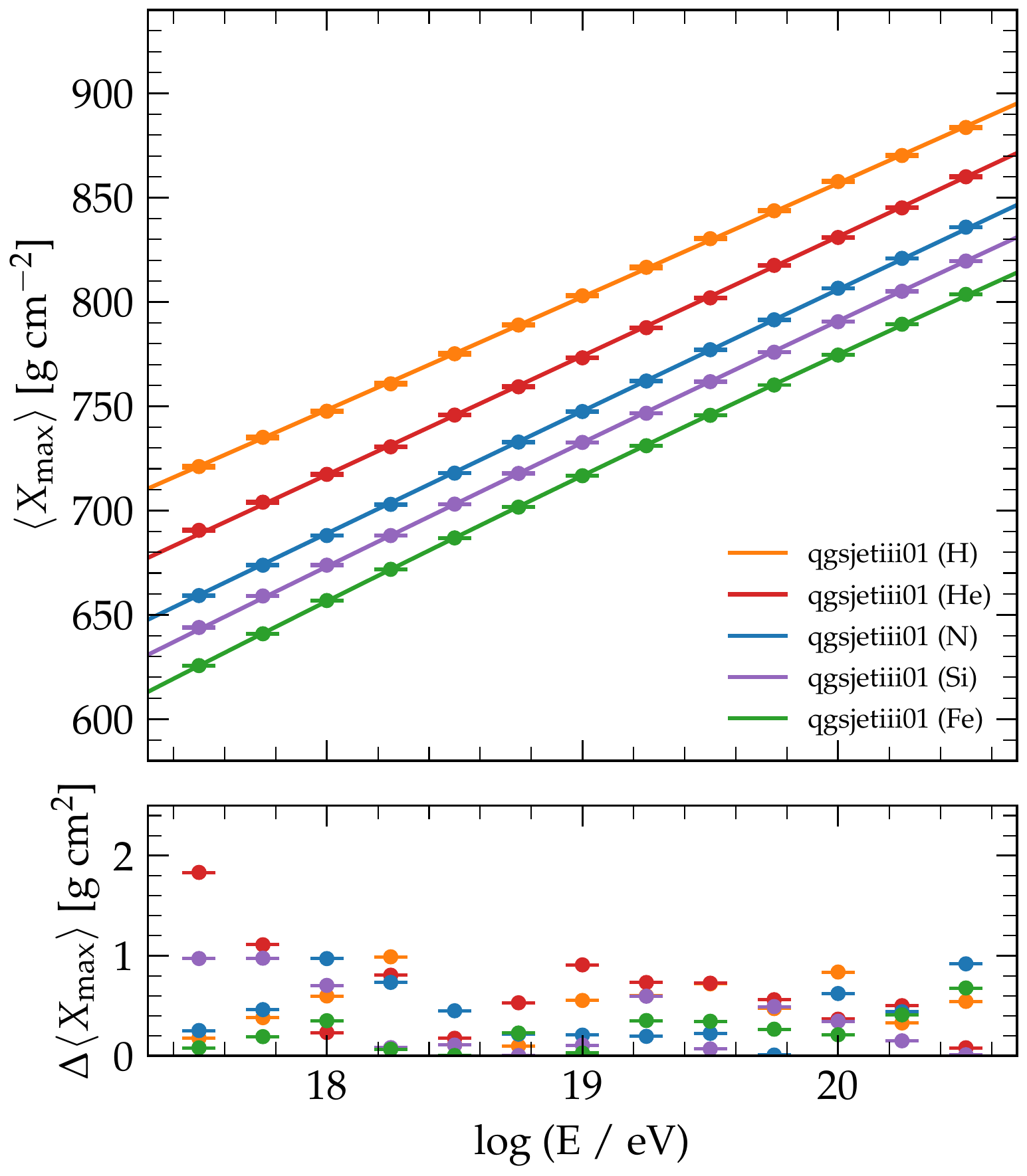}
\caption{Mean depth of shower maximum, $\langle X_{\mathrm{max}}\rangle$, as a function of energy for representative primaries. Upper panels: \conex\ sample means (points) compared to the prediction from the generalized Gumbel parameterization (solid lines), computed via Eq.~\eqref{eq:gumbel_mean}. Lower panels: residuals defined as $\Delta\langle X_{\mathrm{max}}\rangle \equiv \langle X_{\mathrm{max}}\rangle_{\mathrm{MC}}-\langle X_{\mathrm{max}}\rangle_{\mathrm{G}}$. From left to right: \epos, \sibyll, and \qgsjet.}
\label{fig:gumbel_mean}
\end{figure}

\begin{figure}[t]
\centering
\includegraphics[width=0.32\textwidth]{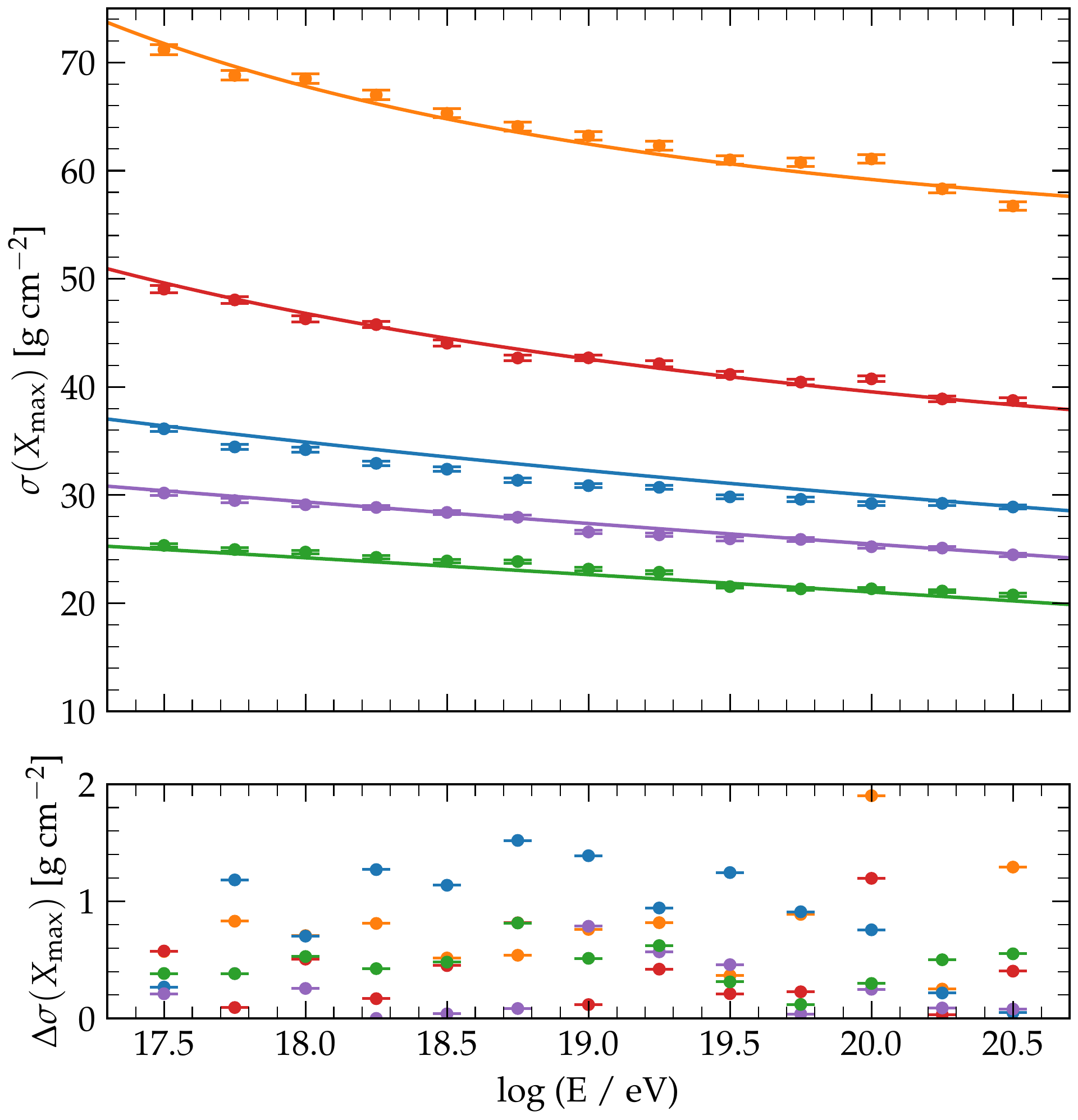}
\includegraphics[width=0.32\textwidth]{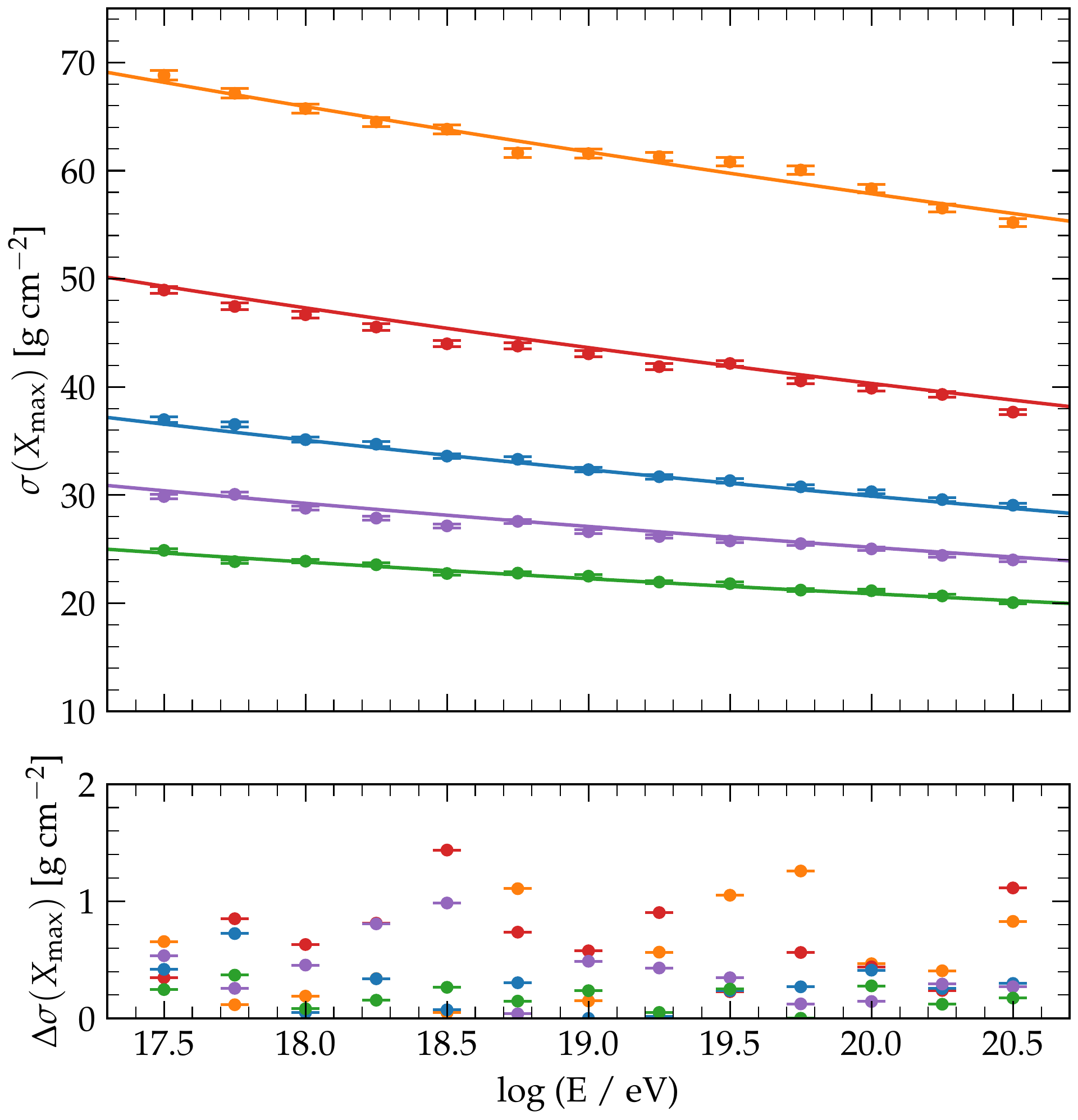}
\includegraphics[width=0.32\textwidth]{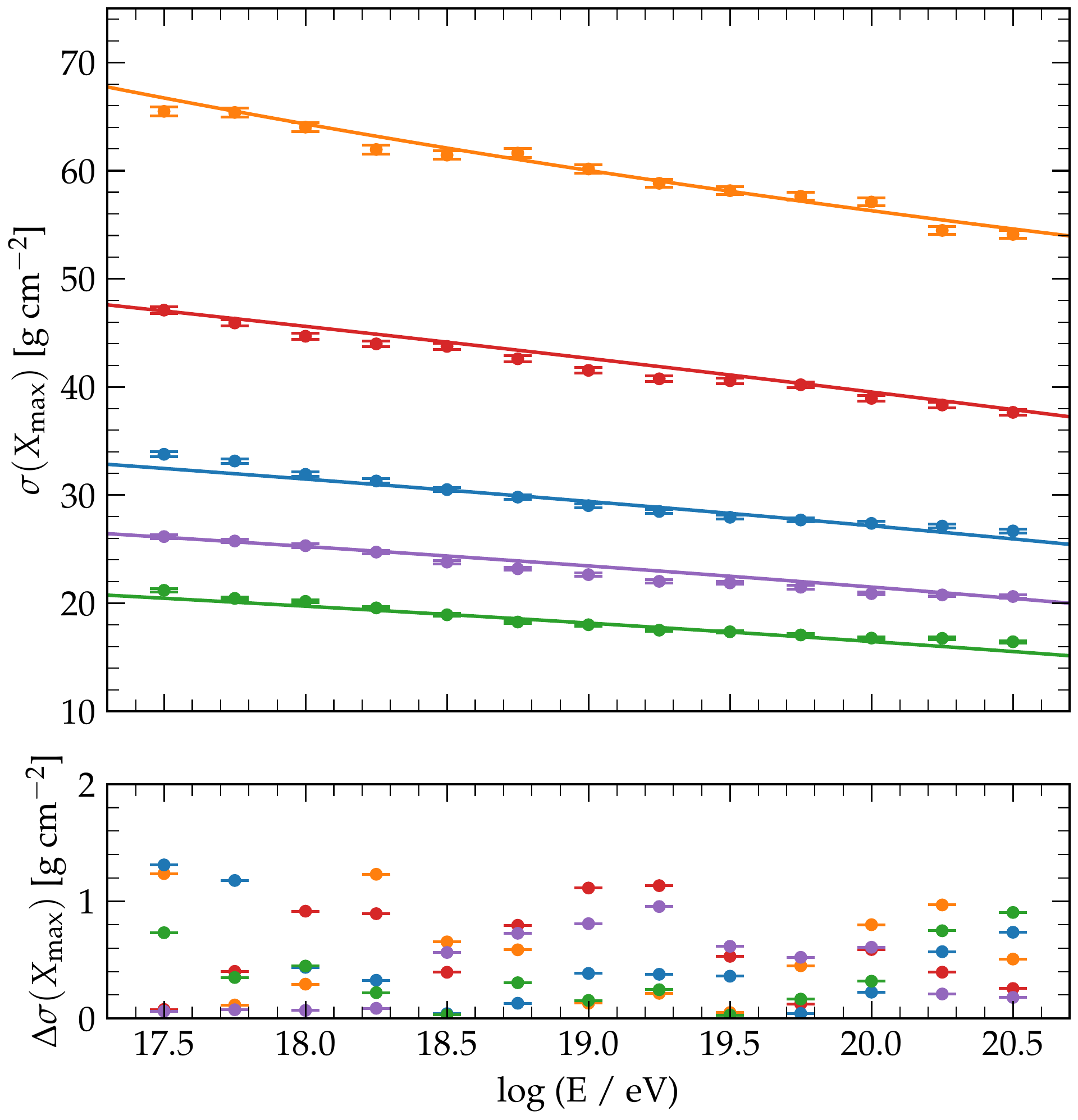}
\caption{Same as Fig.~\ref{fig:gumbel_mean}, but for the $X_{\mathrm{max}}$ standard deviation. Upper panels: $\mathrm{Var}(X_{\mathrm{max}})$ from \conex\ (points) compared to the generalized Gumbel prediction (solid lines), computed via Eq.~\eqref{eq:gumbel_var}. Lower panels: residuals $\Delta \mathrm{Var}(X_{\mathrm{max}})\equiv \mathrm{Var}(X_{\mathrm{max}})_{\mathrm{MC}}-\mathrm{Var}(X_{\mathrm{max}})_{\mathrm{G}}$. From left to right: \epos, \sibyll, and \qgsjet.}
\label{fig:gumbel_sigma}
\end{figure}

For each hadronic interaction model and for each $(E,A)$ point on the simulation grid, we compute:
(i) the sample mean and variance of $X_{\mathrm{max}}$ directly from the \conex\ showers, 
$\langle X_{\mathrm{max}} \rangle_{\mathrm{MC}}$ and $\mathrm{Var}(X_{\mathrm{max}})_{\mathrm{MC}}$;
(ii) the Gumbel-based moments 
$\langle X_{\mathrm{max}} \rangle_{\mathrm{G}}$ and $\mathrm{Var}(X_{\mathrm{max}})_{\mathrm{G}}$ via Eqs.~\eqref{eq:gumbel_mean}–\eqref{eq:gumbel_var}; and
(iii) the moments obtained from the dedicated parameterizations of the first two moments introduced in Sec.~\ref{sec:xmax}.
The corresponding biases are quantified using the definition in Eq.~\eqref{eq:bias}, applied separately to the mean and to the standard deviation.

The comparison is summarized in Fig.s~\ref{fig:gumbel_mean} and \ref{fig:gumbel_sigma}, which shows the residuals between the \conex\ moments and those obtained from the Gumbel representation, together with the residuals from the direct moment parameterizations, as a function of energy for representative primaries. Over the full range of energies and masses considered, the Gumbel-based moments track the \conex\ results at the level of a few $\mathrm{g\,cm^{-2}}$ for the mean and a few percent for the variance, with no significant systematic trend as a function of $E$ or $A$. The size and structure of the residuals are comparable to those obtained with the dedicated moment parameterizations.

We therefore conclude that the generalized Gumbel parameterization provides a faithful and internally consistent description of the $X_{\mathrm{max}}$ distributions: it not only reproduces the detailed histogram shapes and cumulative distributions, but also yields first and second moments that are fully consistent with those from the underlying \conex\ simulations within the accuracy required for combined spectrum and composition analyses.

\section{Conclusions}
\label{sec:conclusions}

We have presented updated parametrizations of the first two moments of the depth of shower maximum, $X_{\rm max}$, and of the full $X_{\rm max}$ distributions for ultra-high-energy cosmic-ray air showers. Using \conex\ simulations for primary nuclei from proton to iron and three modern hadronic interaction models (\sibyll, \epos, and \qgsjet), we derived compact descriptions of $\langle X_{\rm max}\rangle$ and $\sigma^2(X_{\rm max})$ as functions of energy and mass. For the variance, we explored both a second-order polynomial and an exponential model: the polynomial form provides fully analytic forward and inverse mappings between the $X_{\rm max}$ moments and the corresponding $\langle\ln A\rangle$ and ${\rm Var}(\ln A)$, while the exponential case can be inverted numerically following the procedure described in this work.

In addition, we showed that the full $X_{\rm max}$ distributions are well captured by a three-parameter generalized Gumbel function, with parameters obtained from unbinned likelihood fits. The resulting representation reproduces the simulated mean and variance with residuals at the level of a few g\,cm$^{-2}$ and a few (g\,cm$^{-2}$)$^2$, respectively, across the full energy range considered. These parametrizations therefore provide a practical interface between air-shower simulations and composition analyses, enabling efficient forward-folding studies and systematic comparisons across hadronic interaction models.

All the data products associated with this note are made publicly available through a dedicated \texttt{Zenodo} repository~\cite{zenodo} while \conex~simulations are available upon request.

The released material includes the fitting parameters of derived parameterizations, in plain-text format for ease of use.
Each file is provided separately for the three hadronic interaction models (\sibyll, \epos, and \qgsjet). Below we summarize the available files and their content.
\begin{itemize}
\item \texttt{conex\_*.xmax}:
Tables of $X_{\mathrm{max}}$ moments directly extracted from the \conex\ simulations. For each hadronic interaction model and primary nucleus, the file reports the mean $\langle X_{\mathrm{max}} \rangle$ and its statistical uncertainty, as well as the variance $\sigma^2(X_{\mathrm{max}})$ and its uncertainty, as functions of energy. Units are g\,cm$^{-2}$ and g$^2$\,cm$^{-4}$, respectively.

\item \texttt{conex\_*.hist}:
Binned $X_{\mathrm{max}}$ distributions from the \conex\ simulations. Each file covers the range $X_{\mathrm{max}} \in [0,2000]$~g\,cm$^{-2}$, divided into $N=100$ bins of width $\Delta X_{\mathrm{max}} = 20$~g\,cm$^{-2}$. The entries are raw counts per bin (i.e.\ unnormalized histograms).

\item \texttt{xmax\_mean\_*\_EVPS2026.txt}:
Best-fit coefficients of the parametr ization for the mean $X_{\rm max}$ (Eq.~\ref{eq:xmax_mean}). The file format is:
(i) first row: best-fit parameters $(a_0, a_1, b_0, b_1)$;
(ii) second row: corresponding parameter uncertainties;
(iii) third row: $\chi^2$ and the number of degrees of freedom (ndf).

\item \texttt{xmax\_sigma2\_*\_EVPS2026.txt}:
Best-fit coefficients of the parametrization for the variance $\sigma^2(X_{\rm max})$ (Eq.~\ref{eq:xmax_sigma2}). An additional keyword in the filename specifies the chosen functional form for $\sigma^2$ (e.g.\ \texttt{pol2} or \texttt{exp}). The file format is:
(i) first row: best-fit parameters, $(c_0, c_1, c_2, d_0, d_1, d_2)$ for the polynomial model, or $(f_0, f_1, g_0, g_1)$ for the exponential model;
(ii) second row: corresponding parameter uncertainties;
(iii) third row: $\chi^2$ and ndf.

\item \texttt{gumbel\_params\_*\_EVPS2026.txt}:
Best-fit coefficients of the generalized Gumbel parametrization of the full $X_{\rm max}$ distributions (Sec.~\ref{sec:gumbel}). For each hadronic interaction model, the file reports the 21 coefficients entering the parametrizations of the three Gumbel parameters $(\mu,\beta,\lambda)$, denoted as $p_k^{(i)}$ with $i\in\{\mu,\beta,\lambda\}$ and $k=0,\ldots,2$.
\end{itemize}

These resources are intended to facilitate the direct use of the new parametrizations in composition analyses, and to support transparent cross-checks and systematic studies of hadronic-model dependencies.

\section*{Acknowledgements}

We thank Denise Boncioli, Fabio Convegna, Alexey Yushkov, and Michael Unger for fruitful discussions and helpful comments on this work.

\bibliographystyle{elsarticle-num} 
\bibliography{bibliography}

\newpage

\appendix

\section{Comparison with previous HIM versions}
\label{sec:oldhim}

This work updates earlier $X_{\rm max}$ moment parametrizations by adopting the most recent implementations of the HIMs used to generate the reference air-shower predictions.
To ensure continuity with existing UHECR composition analyses---and to facilitate comparisons with results based on earlier parametrizations---we present here a direct, like-for-like comparison with the parametrizations obtained using the previous HIM versions: EPOS-LHC, Sibyll~2.3d, and QGSJet~II-04, which are superseded in this work by EPOS-LHC-R, Sibyll~2.3e, and QGSJet~III-01, respectively.

The comparison is performed using the same functional forms, energy binning, and fitting strategy as in the updated analysis, so that differences can be attributed primarily to the HIM updates.
To this aim, we produced a dedicated set of \conex\ simulations with the earlier HIM versions, using a comparable number of showers and the same energy range as in the updated production.

\begin{figure}[t]
\centering
\includegraphics[width=0.32\textwidth]{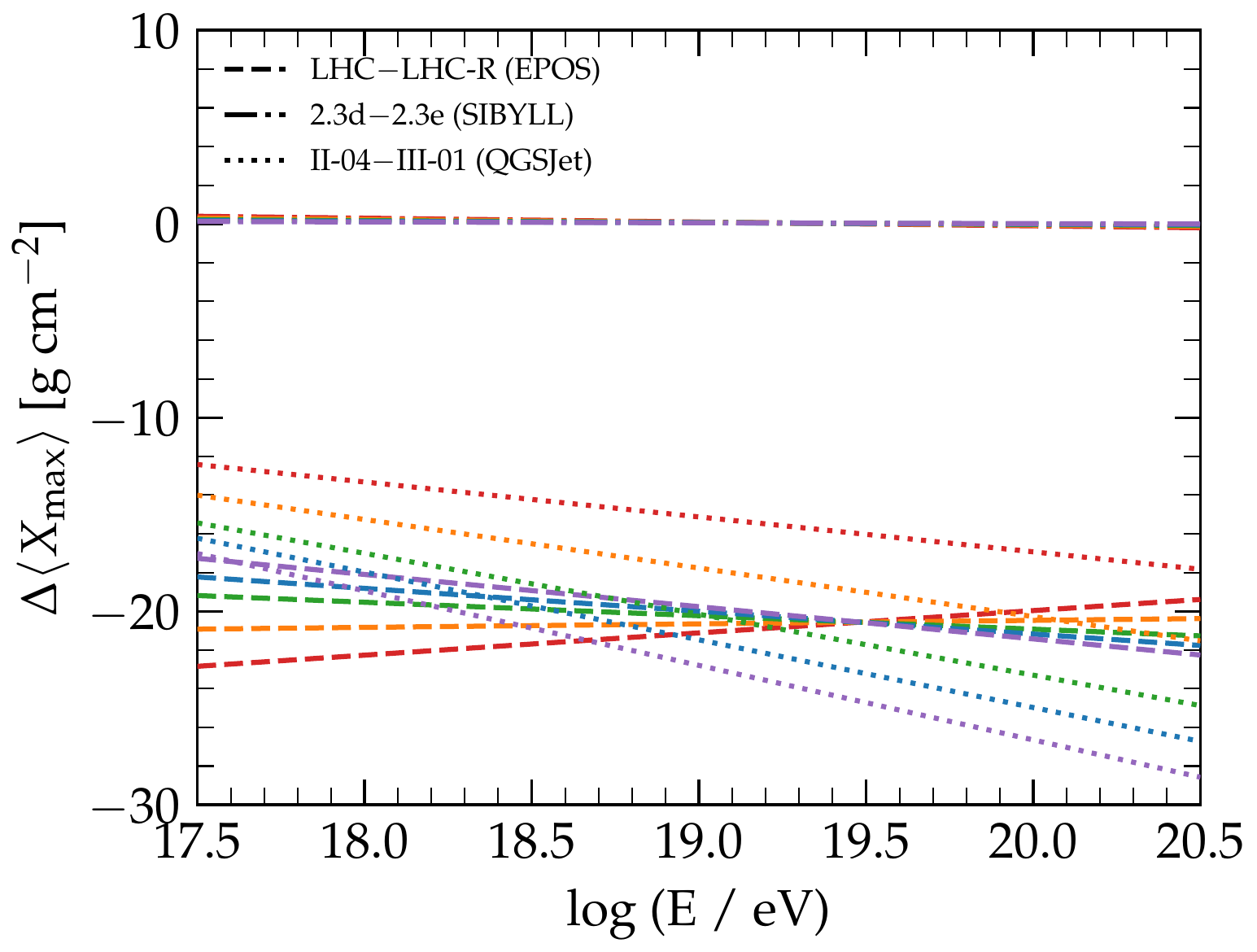}
\includegraphics[width=0.32\textwidth]{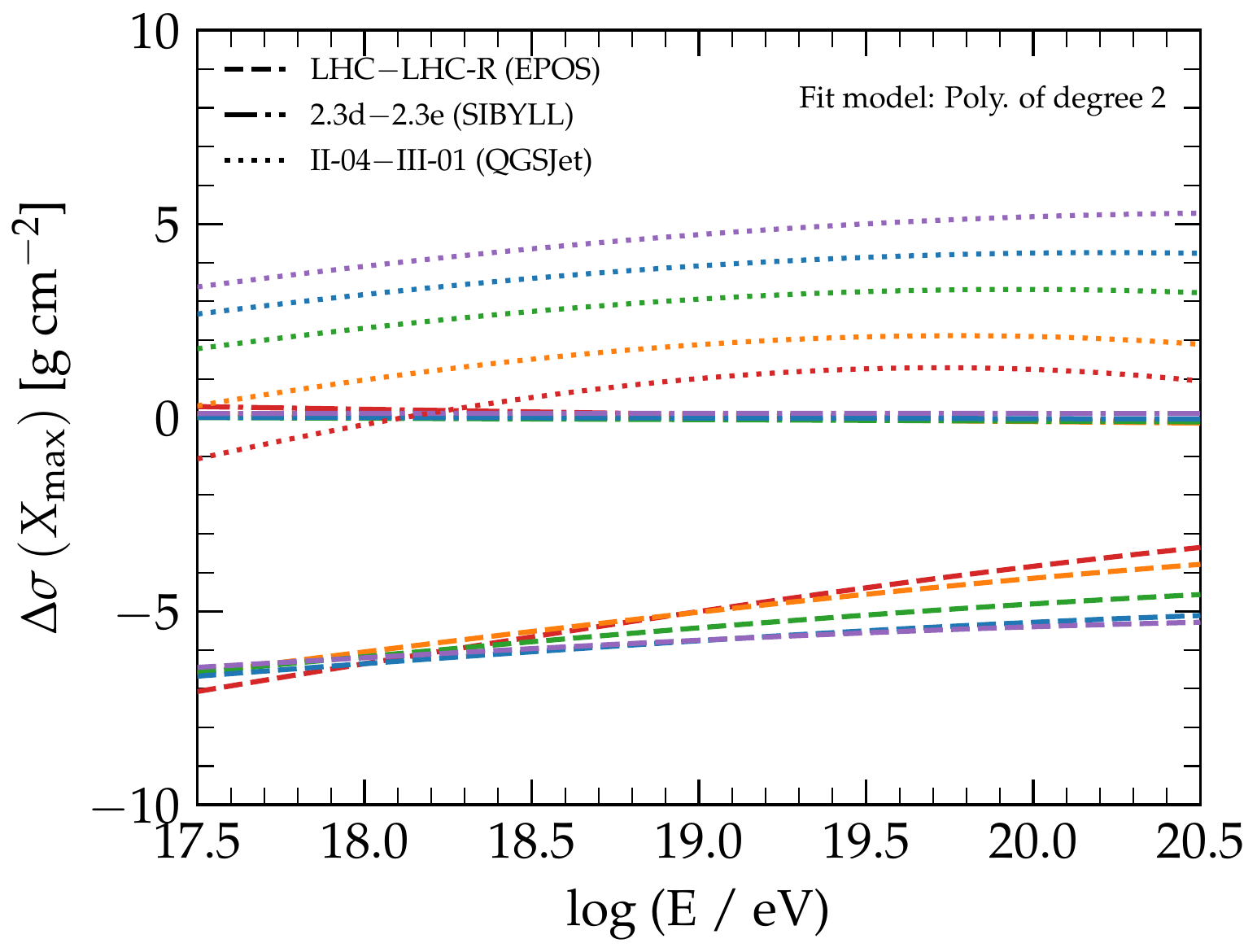}
\includegraphics[width=0.32\textwidth]{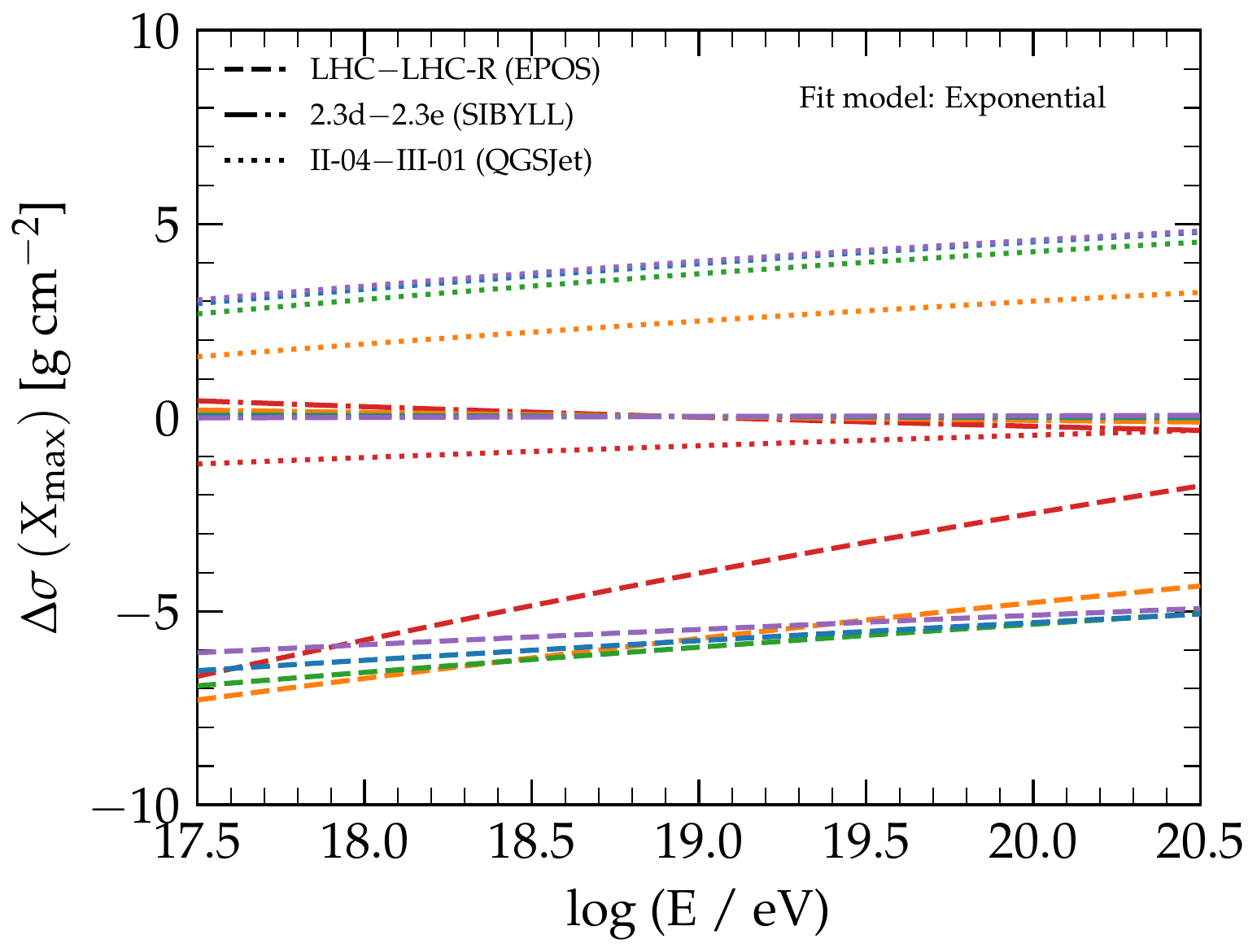}
\caption{Absolute differences between the $X_{\rm max}$ moment parametrizations derived with the updated hadronic interaction models and those obtained with the corresponding earlier versions. Shown are the shifts in $\langle X_{\rm max}\rangle$ (left) and in $\sigma(X_{\rm max})=\sqrt{\sigma^2(X_{\rm max})}$ using the polynomial (pol2) variance model (center) and the exponential (exp) variance model (right), as a function of energy. Curves compare EPOS-LHC-R to EPOS-LHC, Sibyll 2.3e to Sibyll 2.3d, and QGSJet~III-01 to QGSJet~II-04.}
\label{fig:xmax_comparison_new_vs_old}
\end{figure}

The best-fit parameters obtained with the earlier HIM versions are reported in Table~\ref{tab:xmax_moments_params_old}.
A direct comparison between updated and previous parametrizations is shown in Fig.~\ref{fig:xmax_comparison_new_vs_old}, which reports the absolute differences in $\langle X_{\rm max}\rangle$ (left) and in $\sigma(X_{\rm max})$, for both the polynomial (pol2) and exponential (exp) variance models introduced in Section~\ref{sec:xmax}.
These differences quantify the net impact of the hadronic-model updates on the parametrized $X_{\rm max}$ response over the past few years.

Overall, Sibyll exhibits the smallest changes: both the mean and the dispersion remain consistent with the previous parametrization within $\lesssim \mathrm{g\,cm^{-2}}$ over the full energy range considered.
In contrast, EPOS and QGSJet show more pronounced updates.
For EPOS and QGSJet, the updated parametrization predicts systematically deeper $\langle X_{\rm max}\rangle$ by up to ${\cal O}(10)$--${\cal O}(30)\,\mathrm{g\,cm^{-2}}$ (depending on energy), accompanied by a measurable change in the elongation rate at the level of $\sim 10\%$.
For the dispersion, QGSJet tends to yield a smaller $\sigma(X_{\rm max})$, while EPOS shows an increased $\sigma(X_{\rm max})$, with typical differences reaching up to $\sim 5\,\mathrm{g\,cm^{-2}}$.
Such shifts, can propagate to inferences of $\sigma^2(\ln A)$ and therefore remain relevant for precision composition studies.

\newpage

\section{Tables}
\label{sec:tables}

\begin{table}[th]
\begin{center}
\begin{tabular}{|c|c|c|c|c|}
\hline\hline
&  & \epos & \sibyll & \qgsjet \\
\hline\hline
\multirow{5}{*}{$\mu_X$} & $a_0$ & 826.81 $\pm$ 0.11 & 816.44 $\pm$ 0.11 & 804.41 $\pm$ 0.10 \\ 
                                 & $a_1$ & 53.53 $\pm$ 0.12  & 56.93 $\pm$ 0.11    & 55.51 $\pm$ 0.11 \\ 
                                 & $b_0$ & -24.51 $\pm$ 0.03 & -25.82 $\pm$ 0.03  & -21.89 $\pm$ 0.03 \\ 
                                 & $b_1$ & 1.42 $\pm$ 0.04    & 0.75 $\pm$ 0.03     & 0.94 $\pm$ 0.03  \\ 
                                 & $\tilde\chi^2$ & 33.03& 9.97 & 17.23 \\
\hline\hline
\multirow{7}{*}{$\sigma^2_X$ (pol2)} & $c_0$ &  3723 $\pm$ 13  & 3612 $\pm$ 13 & 3431 $\pm$ 12 \\ 
                                 & $c_1$ &  -555 $\pm$ 10& -552 $\pm$ 10 & -497 $\pm$ 9 \\ 
                                 & $c_2$ &   91 $\pm$ 8 & 47 $\pm$ 8 & 93 $\pm$ 7 \\ 
                                 & $d_0$ &  -4.02e-01 $\pm$ 1.1e-03 & -3.78e-01 $\pm$ 1.2e-03 & -4.02e-01 $\pm$ 9.9e-04 \\ 
                                 & $d_1$ &  1.12e-04 $\pm$ 1.7e-04 & 5.49e-04 $\pm$ 1.6e-04 & -9.41e-04 $\pm$ 1.1e-04 \\ 
                                 & $d_2$ &  4.71e-02 $\pm$ 2.7e-04 & 4.08e-02 $\pm$ 2.9e-04 & 4.41e-02 $\pm$ 2.4e-04 \\ 
                                 & $\tilde\chi^2$ & 26.67 & 18.14 & 7.13 \\ 
\hline\hline
\multirow{5}{*}{$\sigma^2_X$ (exp)} & $f_0$ & 8.263 $\pm$ 0.003 & 8.245 $\pm$ 0.003 & 8.239 $\pm$ 0.003 \\
                                 & $f_1$ &  -0.143 $\pm$ 0.003 & -0.146 $\pm$ 0.003 & -0.128 $\pm$ 0.003 \\
                                 & $g_0$ & -0.503 $\pm$ 0.001 & -0.501 $\pm$ 0.001 & -0.589 $\pm$ 0.001 \\
                                 & $g_1$ & -2.30e-04 $\pm$ 1.22e-03 & 3.94e-04 $\pm$ 1.19e-03 & -1.20e-02 $\pm$ 1.18e-03 \\
                                 & $\tilde\chi^2$ & 12.21 & 7.48 & 22.85 \\
\hline\hline
\end{tabular}
\end{center}
\caption{Best-fit parameters for the $\mu_X$ parametrization (Eq.~\ref{eq:xmax_mean}) and for the $\sigma^2(X_{\mathrm{max}})$ parametrizations corresponding to the second-order polynomial model (pol2, Eq.~\ref{eq:xmax_sigma2}) and the exponential model (exp, Eq.~\ref{eq:xmax_sigma2_exp}), obtained from \conex~simulations for the three hadronic interaction models discussed in the main text \epos, \sibyll, and \qgsjet. The last column reports the reduced $\chi^2$.
Parameters $a_0$, $a_1$, $b_0$, and $b_1$ are given in g\,cm$^{-2}$, whereas $c_0$, $c_1$, and $c_2$ are in g$^{2}$\,cm$^{-4}$; all remaining parameters are dimensionless.}
\label{tab:xmax_moments_params}
\end{table}

\begin{table}[th]
\begin{center}
\begin{tabular}{|c|c|c|c|c|}
\hline\hline
&  & \texttt{EPOS LHC} & \texttt{Sibyll 2.3d} & \texttt{QGSJet II-04} \\
\hline\hline
\multirow{5}{*}{$\mu_X$} & $a_0$ & 805.69 $\pm$ 0.10 & 816.54 $\pm$ 0.11 & 789.28 $\pm$ 0.11 \\ 
                                 & $a_1$ &54.68 $\pm$ 0.10 & 56.73 $\pm$ 0.11 & 53.71 $\pm$ 0.11 \\ 
                                 & $b_0$ &  -24.18 $\pm$ 0.03 & -25.83 $\pm$ 0.03 & -23.80 $\pm$ 0.03 \\ 
                                 & $b_1$ &  0.72 $\pm$ 0.03 & 0.79 $\pm$ 0.03 & 0.43 $\pm$ 0.03 \\ 
                                 & $\tilde\chi^2$ &23.12 & 8.51 & 8.36 \\
\hline\hline
\multirow{7}{*}{$\sigma^2_X$ (pol2)} & $c_0$ & 3137 $\pm$ 11 & 3621 $\pm$ 13 & 3550 $\pm$ 13 \\ 
                                 & $c_1$ & -367 $\pm$ 8 & -569 $\pm$ 10 & -416 $\pm$ 9 \\ 
                                 & $c_2$ & 65 $\pm$ 7 & 49 $\pm$ 8 & 32 $\pm$ 7 \\
                                 & $d_0$ & -4.35e-01 $\pm$ 8.5e-04 & -3.80e-01 $\pm$ 1.2e-03 & -3.77e-01 $\pm$ 1.2e-03 \\
                                 & $d_1$ & -7.19e-04 $\pm$ 1.1e-04 & 7.19e-04 $\pm$ 1.6e-04 & 1.01e-03 $\pm$ 1.7e-04 \\ 
                                 & $d_2$ & 5.22e-02 $\pm$ 2.1e-04 & 4.14e-02 $\pm$ 2.9e-04 & 4.12e-02 $\pm$ 3.0e-04 \\ 
                                 & $\tilde\chi^2$ & 24.83 & 19.26 & 9.13 \\
\hline\hline
\multirow{5}{*}{$\sigma^2_X$ (exp)} & $f_0$ & 8.129 $\pm$ 0.003 & 8.245 $\pm$ 0.003 & 8.215 $\pm$ 0.003 \\
                                 & $f_1$ & -0.097 $\pm$ 0.003 & -0.154 $\pm$ 0.003 & -0.120 $\pm$ 0.003 \\
                                 & $g_0$ & -0.608 $\pm$ 0.001 & -0.500 $\pm$ 0.001 & -0.486 $\pm$ 0.001 \\
                                 & $g_1$ & -1.21e-02 $\pm$ 1.23e-03&  2.95e-03 $\pm$ 1.19e-03 & 6.60e-03 $\pm$ 1.20e-03 \\
                                 & $\tilde\chi^2$ & 7.22 & 6.46 & 4.04 \\
\hline\hline
\end{tabular}
\end{center}
\caption{Best-fit parameters for the $\mu_X$ parametrization (Eq.~\ref{eq:xmax_mean}) and for the $\sigma^2(X_{\mathrm{max}})$ parametrizations corresponding to the second-order polynomial model (pol2, Eq.~\ref{eq:xmax_sigma2}) and the exponential model (exp, Eq.~\ref{eq:xmax_sigma2_exp}), obtained from \conex~simulations for the earlier verion of the three hadronic interaction models discussed in the main text: \texttt{EPOS LHC}, \texttt{Sibyll 2.3e}, and \texttt{QGSJet II-04}. 
The last column reports the reduced $\chi^2$.
Parameters $a_0$, $a_1$, $b_0$, and $b_1$ are given in g\,cm$^{-2}$, whereas $c_0$, $c_1$, and $c_2$ are in g$^{2}$\,cm$^{-4}$; all remaining parameters are dimensionless.}
\label{tab:xmax_moments_params_old}
\end{table}

\begin{table}[htp]
\begin{center}
\begin{tabular}{|c|c|c|c|c|c|}
\hline\hline
HIM & H & He & N & Si & Fe \\
\hline
\epos   & 2.05 & 1.03 & 1.05 & 1.44 & 1.45 \\ 
\sibyll & 1.46 & 0.77 & 0.52 & 0.53 & 0.75 \\ 
\qgsjet & 1.83 & 0.61 & 1.02 & 0.66 & 0.67 \\ 
\hline\hline
\end{tabular}
\end{center}
\caption{Average residuals between the \conex~simulations and the analytical parametrizations for $\mu_X$, computed for each primary species and hadronic 
interaction model in the energy range $E \gtrsim 10^{17.5}$~eV. Values are expressed in g\,cm$^{-2}$.}
\label{tab:xmax_mean_residuals}
\end{table}

\begin{table}[htp]
\begin{center}
\begin{tabular}{|c|c|c|c|c|c|c|}
\hline\hline
HIM &  & H & He & N & Si & Fe \\
\hline	
\multirow{2}{*}{\epos}   & pol2 & 2.19 & 1.78 & 0.38 & 0.91 & 0.38 \\
                         & exp & 1.31 & 0.92 & 0.68 & 0.28 & 0.40 \\
\multirow{2}{*}{\sibyll} & pol2 & 1.67 & 1.75 & 0.43 & 0.35 & 0.18 \\
                         & exp & 0.63 & 0.64 & 0.77 & 0.22 & 0.24 \\
\multirow{2}{*}{\qgsjet} & pol2 & 1.03 & 0.95 & 0.27 & 0.17 & 0.14 \\
                         & exp & 1.60 & 1.01 & 1.24 & 0.24 & 0.53 \\
\hline\hline
\end{tabular}
\end{center}
\caption{Average residuals between the \conex~simulations and the analytical parametrizations for $\sigma(X_{\mathrm{max}})$, computed for each primary species and hadronic 
interaction model in the energy range $E \gtrsim 10^{17.5}$~eV and for the two parametrizations proposed in Eqs.~\eqref{eq:xmax_sigma2} (2nd) and~\eqref{eq:xmax_sigma2_exp} (exp). Values are expressed in g\,cm$^{-2}$.}
\label{tab:xmax_sigma_residuals}
\end{table}

{\def\arraystretch{1.5}
\begin{table}[htp]
\begin{center}
\begin{tabular}{|c|c|c|c|c|c|c|c|c|c|c|}
\hline\hline
& \multicolumn{3}{|c|}{\epos} & \multicolumn{3}{|c|}{\sibyll} & \multicolumn{3}{|c|}{\qgsjet} \\
\hline
$p_0^\mu$ & 796.26 & -12.23 & -1.55          & 786.20 & -15.99 & -1.03     & 770.84 & -8.69 & -1.50  \\
$p_1^\mu$& 58.67 & -0.85 & 0.29                & 59.62 & 0.22 & 0.02          & 57.22 & 0.68 & -0.05  \\
$p_2^\mu$ & -1.418 & -0.172 & -0.009         & -1.132 & 0.472 & -0.116    & -0.084 & 0.368 & -0.153  \\
$p_0^\sigma$ & 43.413 & -0.243 & -0.830   & 40.417 & -1.956 & -0.427  & 33.632 & 3.985 & -1.276  \\
$p_1^\sigma$& 3.348 & -1.731 & 0.067       & -1.207 & -0.238 & 0.094    & -0.289 & -3.100 & 0.427  \\
$p_0^\lambda$ & 0.857 & 0.406 & -0.025    & 0.793 & 0.231 & 0.003      & 0.651 & 0.238 & 0.087  \\
$p_1^\lambda$& 0.161 & 0.093 & -0.037     & 0.036 & 0.025 & -0.001   & 0.047 & -0.026 & -0.032  \\
\hline\hline
\end{tabular}
\end{center}
\caption{Best-fit coefficients of the generalized Gumbel parameterization of the $X_{\mathrm{max}}$ distributions for the three hadronic interaction models \epos, \sibyll, and \qgsjet, obtained from a global unbinned maximum-likelihood fit to \conex\ simulations. For each model, the three entries in a row correspond to the coefficients of the quadratic dependence on $Y=\ln A$ defining $p_j(Y)=a_j+b_j Y+c_j Y^2$ (see Eqs.~\eqref{eq:mu}--\eqref{eq:lambda}).}\label{tab:gumbel_parameter_fit}
\end{table}}

\end{document}